\documentclass{aa}  

\def\apjl{ApJ}
\def\apjs{ApJS}
\def\ao{Appl.~Opt.}

\def\aap{A\&A}

\usepackage{newtxtext,newtxmath}

\usepackage[T1]{fontenc}
\usepackage{ae,aecompl}
\usepackage{xfrac}
\usepackage{nicefrac}

\usepackage{graphicx}	
\usepackage{amsmath}	
\usepackage{amssymb}	
\usepackage{multirow}
\makeatletter
\newcommand\footnoteref[1]{\protected@xdef\@thefnmark{\ref{#1}}\@footnotemark}
\makeatother
\usepackage{threeparttable}
\usepackage{color}
\usepackage{txfonts}
\usepackage{url}

\def\kms{\mbox{${\rm km}\:{\rm s}^{-1}$}}
\def\arcsec{\hbox{$^{\prime\prime}$}}

\def\degree{\mbox{$^{\circ}$}}

\usepackage{mathtools}
\usepackage{appendix}
\usepackage{adjustbox}

\DeclarePairedDelimiter\abs{\lvert}{\rvert}
\bibpunct{(}{)}{;}{a}{}{,} 
\begin{document}
\titlerunning{Thick-disk origin in the Fornax cluster}
\title{The Fornax\,3D project: Thick disks in a cluster environment}


\author{F. Pinna\inst{1}\fnmsep\inst{2}\fnmsep\thanks{E-mail: fpinna@iac.es},
J. Falc\'on-Barroso\inst{1}\fnmsep\inst{2},
M. Martig\inst{3},
L. Coccato\inst{4},
E. M. Corsini\inst{5}\fnmsep\inst{6},
P.T. de Zeeuw\inst{7}\fnmsep\inst{8},
D.A. Gadotti\inst{4},
E. Iodice\inst{9},
R. Leaman\inst{10},
M. Lyubenova\inst{4},
I. Mart\'in-Navarro\inst{11}\fnmsep\inst{10}
L. Morelli\inst{12}\fnmsep\inst{6},
M. Sarzi\inst{13},
G. van de Ven\inst{14}\fnmsep\inst{4},
S. Viaene\inst{15}\fnmsep\inst{16},
R.M. McDermid\inst{17}\fnmsep\inst{18}
}

\authorrunning{F. Pinna et al.}

\institute{
{Instituto de Astrof\'isica de Canarias, Calle Via L\'actea s/n, E-38200 La Laguna, Tenerife, Spain}
\and
{Depto. Astrof\'isica, Universidad de La Laguna, Calle Astrof\'isico Francisco S\'anchez s/n, E-38206 La Laguna, Tenerife, Spain}
\and
{Astrophysics Research Institute, Liverpool John Moores University, 146 Brownlow Hill, Liverpool L3 5RF, UK}
\and
{European Southern Observatory, Karl-Schwarzschild-Strasse 2, 85748 Garching bei Muenchen, Germany}
\and 
{Dipartimento di Fisica e Astronomia `G. Galilei', Universit\`a di
Padova, vicolo dell'Osservatorio 3, I-35122 Padova, Italy}
\and 
{INAF--Osservatorio Astronomico di Padova, vicolo dell'Osservatorio 5, I-35122 Padova, Italy}
\and
{Sterrewacht Leiden, Leiden University, Postbus 9513, 2300 RA Leiden, The Netherlands}
\and
{Max-Planck-Institut fuer extraterrestrische Physik, Giessenbachstrasse, 85741 Garching bei Muenchen, Germany}
\and
{INAF--Osservatorio Astronomico di Capodimonte, via Moiariello 16, I-80131 Napoli, Italy}
\and
{Max-Planck Institut fuer Astronomie, Konigstuhl 17, D-69117 Heidelberg, Germany}
\and
{University of California Observatories, 1156 High Street, CA-95064, Santa Cruz, USA}
\and
{Instituto de Astronom\'ia y Ciencias Planetarias, Universidad de Atacama, Copiap\'o, Chile}
\and
{Armagh Observatory and Planetarium, College Hill, Armagh, BT61 9DG, UK}
\and
{Department of Astrophysics, University of Vienna, T\"urkenschanzstrasse 17, 1180 Wien, Austria}
\and
{Sterrenkundig Observatorium, Universiteit Gent, Krijgslaan 281, B-9000 Gent, Belgium}
\and
{Centre for Astrophysics Research, University of Hertfordshire, College Lane, Hatfield AL10 9AB, UK}
\and
{Department of Physics and Astronomy, Macquarie University, Sydney, NSW 2109, Australia}
\and
{Australian Astronomical Observatory, PO Box 915, Sydney, NSW 1670, Australia}
}

\date{Received XXX; accepted YYY}

 
  \abstract
   {We used deep MUSE observations to perform
a stellar-kinematic and population analysis of FCC\,153 and FCC\,177, two edge-on S0 galaxies in the Fornax cluster.
The geometrical definition of the different structural components of these two galaxies allows us to describe the nature of their thick disks.
These are both old, relatively metal poor and [Mg/Fe]-enhanced, and their star formation history (SFH) reveals a  minor younger component whose chemical properties suggest its later accretion.
Moreover, the outer regions of these geometrically defined thick disks show higher values of metallicity and lower values of [Mg/Fe]. These stars probably formed in the thin-disk region and they were dynamically heated to form the flares present in these two galaxies.
We propose different formation scenarios for the three populations of these thick disks: in-situ formation, accretion and disk heating.
A clear distinction in age is found between the metal poor and [Mg/Fe]-enhanced thick disks (old, $\sim 12-13$\,Gyr), and the metal rich and less [Mg/Fe]-enhanced thin disks (young, $\sim 4-5$\,Gyr).
These two galaxies show signs of relatively recent star formation in their thin disks and nuclear regions. While the thin disks show more continuous SFHs, the nuclei display a rather bursty SFH.
These two galaxies are located outside of the densest region of the Fornax cluster where FCC\,170 resides. 
This other edge-on S0 galaxy was studied by \citet{Pinna2019}. We compare and discuss our results with this previous study.
The differences between these three galaxies, at different distances from the cluster center, suggest that the environment can have a strong effect on the galaxy evolutionary path. \looseness-2
}

\keywords{
galaxies: kinematics and dynamics -- galaxies: evolution -- galaxies: elliptical and lenticular, cD -- galaxies: structure -- galaxies: formation -- galaxies: individual: IC\,1963, NGC\,1380A}

\maketitle

\section{Introduction}  \label{sec:intro}
\begin{table*}[!h]
\centering
\caption{Relevant properties of the three galaxies FCC\,153 and FCC\,177, analyzed here, and FCC\,170, analyzed in Paper~I.}
\begin{adjustbox}{max width=\textwidth}

\begin{tabular}{||l|c|c|c|c|c|c|c|c|c|c|c|r||} 
\hline\hline

FCC$^{\left(1\right)}$ & Alternative & D$^{\left(2\right)}$ & \multicolumn{2}{|c|}{$D_\text{c}^{\left(3\right)}$} & \multicolumn{2}{|c|}{$R_\text{e}^{\left(4\right)}$} & ${V_{\text{c,max}}}^{\left(5\right)}$ & \multirow{1}{*}{\textit{M}$_{{\ast}}^{\left(6\right)}$} & \multicolumn{2}{|c|}{${z_{\text{c1}}}^{\left(7\right)}$}& \multicolumn{2}{|r||}{${r_{\text{NSC}}}^{\left(8\right)}$}\\
\cline{3-13}
name & names & (Mpc) & (\degree) & (Mpc) & (\arcsec) & (kpc) & (\kms) & ($10^9$M$_{\odot}$)  & (\arcsec) & (kpc)& (\arcsec) & (kpc)\\
\hline\hline
\multirow{2}{*}{153} & IC\,1963 & \multirow{2}{*}{20.8} & \multirow{2}{*}{1.17} & \multirow{2}{*}{$0.45$} & \multirow{2}{*}{11.4} & \multirow{2}{*}{$1.15$} & \multirow{2}{*}{$165$} & \multirow{2}{*}{$7.6$}& \multirow{2}{*}{5.1}
&{\multirow{2}{*}{$0.5$}}& \multirow{2}{*}{$0.2$}
& \multicolumn{1}{|r||}{\multirow{2}{*}{$0.02$}}\\
& IC\,0335 & &&&&&&&&\multicolumn{1}{|c|}{}&&\multicolumn{1}{|r||}{} \\
\hline
\multirow{2}{*}{177} & NGC\,1380A & \multirow{2}{*}{20.0} & \multirow{2}{*}{0.79} & \multirow{2}{*}{$0.30$} & \multirow{2}{*}{15.0} & \multirow{2}{*}{$1.45$} & \multirow{2}{*}{$120$} &\multirow{2}{*}{$8.5$}& \multirow{2}{*}{8.4} & {\multirow{2}{*}{$0.8$}}& \multirow{2}{*}{$0.4$} & \multicolumn{1}{|r||}{\multirow{2}{*}{$0.04$}}\\
& PGC\,13335 & &&&&&&&&\multicolumn{1}{|c|}{}&&\multicolumn{1}{|r||}{} \\
\hline
\multirow{2}{*}{170} & NGC\,1381 & \multirow{2}{*}{21.9} & \multirow{2}{*}{0.42} & \multirow{2}{*}{$0.16$} & \multirow{2}{*}{13.8} & \multirow{2}{*}{$1.47$} & \multirow{2}{*}{$280$} & \multirow{2}{*}{$22.5$}& \multirow{2}{*}{10.8} & {\multirow{2}{*}{$1.1$}}& \multirow{2}{*}{} & \multicolumn{1}{|r||}{\multirow{2}{*}{ }}\\
& PGC\,13321 & &&&&&&&&\multicolumn{1}{|c|}{}&&\multicolumn{1}{|r||}{} \\
\hline
\hline
\end{tabular}
\end{adjustbox}
\label{chap4/table:gal_prop}
\begin{tablenotes}
\item {\footnotesize Notes. (1) Galaxy name from the Fornax Cluster Catalog \citep{Ferguson1989a}.
(2) Distance from \citet{Blakeslee2009}.
(3) Projected distance from the cluster central galaxy FCC\,213 \citep{Iodice2019}.
(4) Effective radius in the $B$ band \citep{Ferguson1989b}.
(5) Maximum circular velocity, from \citet{Bedregal2006}.
(6) Stellar mass as estimated from the $i$ band \citep{Iodice2019}.
(7) Lower limit of the region where the thick disk dominates, according to the photometrical decomposition by \citet{Comeron2018}.
(8) Mean radius of the region dominated by the NSC (from Fig.~3 in \citealt{Turner2012}).
}
\end{tablenotes}

\end{table*}
The definition of what the thick disk is, as a structural component, has not been unique in the literature.
Different ways of defining thick disks have been used in different studies, especially in the Milky Way, leading to slightly different properties. 
The chemical definition, widely used for the Milky Way, is based on the $\alpha$-enhancement of thick disks, which form a different sequence in the [$\alpha$/Fe]-[Fe/H] diagram with respect to the thin disk \citep[e.g.,][]{Fuhrmann1998, Fuhrmann2011,Bovy2012}.
The geometrical (or morphological) definition is based on a photometrical decomposition \citep[e.g.,][]{Yoachim2006,Comeron2012,Comeron2018}. 
It is mostly used in external galaxies where, unless they are very nearby \citep[e.g.,][]{Seth2005,Tikhonov2005,Rejkuba2009}, we cannot resolve the individual stars.
It is also possible to use kinematics to perform a decomposition \citep[e.g.,][]{Morrison1990},
while a decomposition based on age has also been proposed \citep[e.g.,][]{Haywood2013}. \looseness-2

Recent works on stellar populations of thick disks in external galaxies have globally shown their diversity. 
Age and metallicity in wide ranges were associated by \citet{Yoachim2008b} and \citet{Kasparova2016} to thick disks in their samples of late-type and S0-a galaxies, respectively.
The thick-disk relative properties compared to the respective thin disks display also a certain variety. 
On the one hand, a clear distinction in age between the thick and the thin disk has been shown in most of the analyzed galaxies (see also \citealt{Comeron2015}).
On the other hand, two very old S0 galaxies were found in clusters by \citet{Comeron2016} and \citet{Pinna2019} (hereafter Paper~I, this was the first paper of the series on thick disks within the Fornax\,3D project). The thin and thick disks of these two galaxies showed approximately the same age, but a clear difference in their chemical properties unveiled the different origin and evolution of the two disks. \looseness-2

This variety of thick disks points to different origins among galaxies with different structures and in different environments.
The most invoked scenarios support either an internal or an external origin.
Thick disks could have been formed in situ from dynamically hot turbulent gas at high redshift, during the phase of multiple gas-rich mergers \citep{Brook2004}.
Alternatively, its stars might have come from a preexisting disk, dynamically cooler and thinner, which would have been dynamically heated \citep[e.g.,][]{Bournaud2009,Helmi2018}. This disk thickening could have happened fast at high redshift, with minor mergers being the most invoked process responsible for it \citep[e.g.,][]{Quinn1993, Kazantzidis2008}.
Another possibility is that thick-disk stars were formed ex situ and accreted later \citep{Abadi2003}.
In their observational analyses, \citet{Comeron2015} preferred a ''born hot'' thick disk for ESO\,533-4 and \citet{Comeron2016} proposed a ''monolithic collapse'' for the formation of both the thin and the thick disks in ESO\,243-49.
On the other hand, \citet{Yoachim2005,Yoachim2006,Yoachim2008a,Yoachim2008b} strongly supported accretion as formation scenario for the thick disks of their sample.
Nonetheless, different mechanisms could follow one another and be responsible for the formation of the thick disks that we observe today.
A combination of both stars formed in situ and ex situ was proposed in Paper~I to explain the complexity of the thick disk observed in FCC\,170. \looseness-2

Environment, closely connected to the global history of a galaxy, could have important consequences also in the nature of its thick disk \citep{Kasparova2016,Katkov2019}.
The galaxy infall into a cluster could initially accelerate the star formation, which could also be quenched by stripping (or starvation) processes, when approaching to the cluster core \citep[e.g.,][]{Fujita1999}. 
As proposed in Paper~I for FCC\,170, this evolution could also happen during the "preprocessing" phase in galaxy groups before entering the cluster \citep{Haines2015,Sarron2019}.
At the same time, mergers could contribute with new populations \citep{Abadi2003} or create flares in the disks \citep{Bournaud2009,Qu2011}.
The relation between thick-disk origin and galaxy formation also stresses the importance of thick-disk studies to probe the proposed evolution scenarios in different environments. \looseness-2

Considering the small sample of galaxies whose thick disks have been exhaustively analyzed, this work aims to contribute the stellar kinematics and populations of two more galaxies. 
In this second paper of the series, we perform the same analysis done as in Paper~I for two more edge-on S0 galaxies in the Fornax cluster.
This paper is outlined as follows. 
Section~\ref{sec:sample} describes the sample. Sections~\ref{sec:obs} and \ref{sec:methods} characterize briefly the data set and the methods, leading to the results presented in \S~\ref{sec:results}. In \S~\ref{sec:discussion} we discuss our results, also in the more general context of the Fornax cluster. Section~\ref{sec:conclusions} summarizes our conclusions. \looseness-2

\section{The sample}
\label{sec:sample}
\begin{figure*}[!h]
\begin{minipage}[H!]{1\textwidth}
\centering
	\includegraphics[scale=6.5]{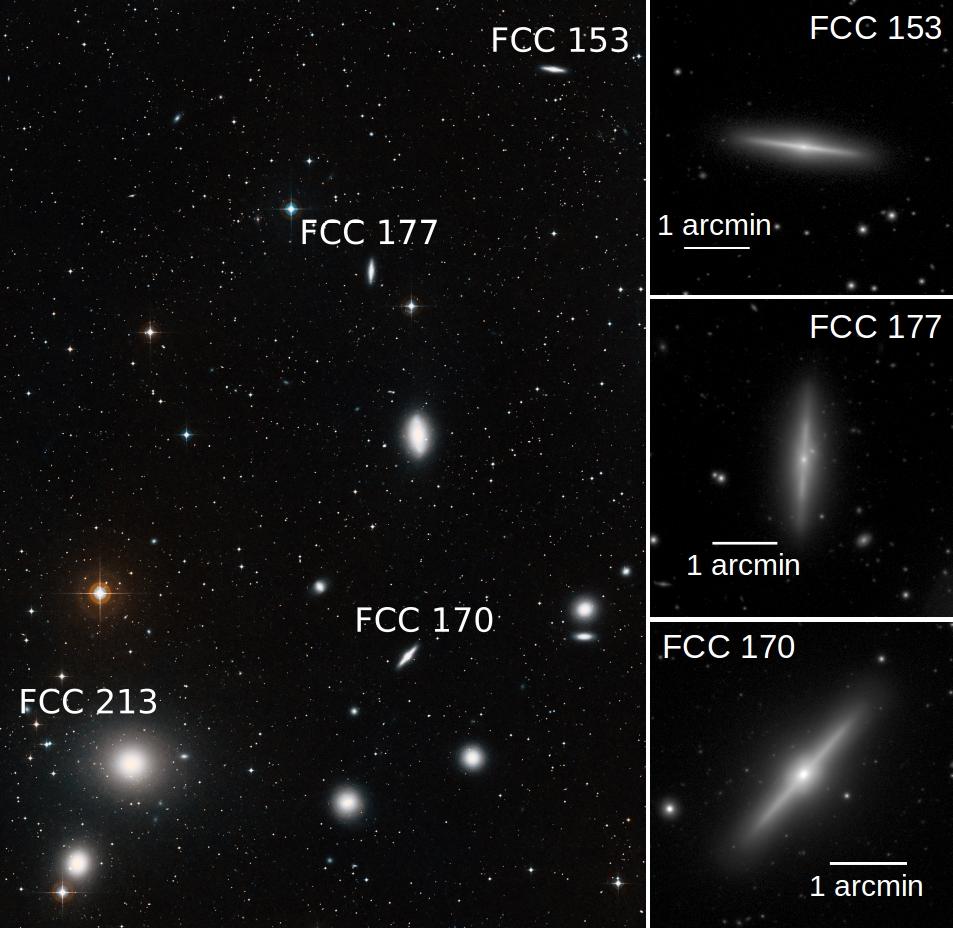}
\end{minipage}
\caption{
Left: clipping of the wide-field view of the Fornax Galaxy Cluster
 (credit: ESO and Digitized Sky Survey 2, acknowledgment: Davide De Martin).
The three galaxies analyzed in Paper~I and this paper are indicated with their names.
FCC\,213 is the central galaxy of the cluster.
Right: $r$-band images from the Fornax Deep Survey \citep{Iodice2019}. From top to bottom: FCC\,153, FCC\,177 and FCC\,170. A scale bar is indicated in each image.
}
\label{chap4/fig:fornax}
\end{figure*}
The galaxies analyzed in this paper are the two S0 galaxies FCC\,153 and FCC\,177. They were selected because they are observed edge-on and this property is crucial to allow a geometrical decomposition between the thin and the thick disk.
Similar to FCC\,170 (studied in Paper~I), they belong to the Fornax cluster, a dense and dynamically relaxed nearby system \citep[e.g.,][]{Sarzi2018}.
Thus, the properties described in this work could be considered as representative of S0 galaxies in clusters. Their evolution histories could test the role of environment in shaping lenticular galaxies.
Furthermore, their different (projected) positions in the cluster, as indicated in Fig.~\ref{chap4/fig:fornax}, make this work an excellent opportunity to discuss the galaxy evolution in different locations of the same cluster.
The most relevant properties of FCC\,153 and FCC\,177 are summed up in Table~\ref{chap4/table:gal_prop} (where we have also added FCC\,170 for completeness) and their images in the $r$ band are shown in Fig.~\ref{chap4/fig:fornax}.
Their distances are such that 10\,arcsec on the sky correspond to about 1.01 and 0.97\,kpc, respectively. 
When it comes to the thick-to-thin luminosity and mass ratios, we expect FCC\,153 to behave as a massive galaxy and to be thin-disk dominated ($V_{\text{c,max}} \sim 165$\,\kms, see Table~\ref{chap4/table:gal_prop}), according to the criterion by \citet{Yoachim2006}. Because of its lower maximum circular velocity ($V_{\text{c,max}} \sim 120$\,\kms), in spite of a stellar mass slightly larger than FCC\,153 (see Table~\ref{chap4/table:gal_prop}), FCC\,177 is expected to be an intermediate case between thin and thick-disk dominated. \looseness-2

The midplane kinematics as extracted by \citet{Donofrio1995} and \citet{Bedregal2006} do not show, in either galaxy, any other prominent structure than an edge-on disk, which starts to dominate from very close to the galaxy center.
In agreement with this,
\citet{Johnston2012} used a single disk component to fit FCC\,153 in their spectroscopic bulge-disk decomposition.
\citet{Lutticke2000} classified the bulge of FCC\,177 as a boxy bulge, but the one in FCC\,153 as an elliptical bulge.
\citet{Comeron2018} performed a photometric decomposition using two disk components and a small central mass concentration for both galaxies in our sample. They calculated the average height $z_{\text{c1}}$ above which most of the light is emitted by the thick disk.
Its values, in Table~\ref{chap4/table:gal_prop}, indicate a thicker thin disk for FCC\,177 than in FCC\,153.
\citet{Turner2012} identified nuclear star clusters (NSCs) in the center of numerous galaxies of the Fornax cluster, including FCC\,153 and FCC\,177.
Their NSCs were found to be bluer than their hosts, suggesting the predominance of younger populations of stars.
According to their surface brightness analysis, the nucleus dominates over the host up to a mean radius of about 0.4\,arcsec in FCC\,177 and 0.2\,arcsec in FCC\,153. Nevertheless, in the latter galaxy, the surface brightness of the nucleus is not much larger than that of the host, even in the center. \looseness-2

\section{Observations and data reduction} \label{sec:obs}
Deep high-quality integral-field data from the Fornax\,3D survey \citep{Sarzi2018} were used for the present work. They were collected at the UT\,4 at the Very Large Telescope (VLT), by means of the Multi Unit Spectroscopic Explorer (MUSE) \citep{Bacon2010}. 
Observations were performed in the same way as for FCC\,170. They follow the description in Paper~I and are detailed in \citet{Sarzi2018}.
Two MUSE pointings were used for each one of the two galaxies in our sample. The position of these pointings is indicated in Fig.~A.1 of \citet{Sarzi2018}.
On-source exposure times for the central pointings were the same as for FCC\,170 (five exposures of 720\,s), and they were 1.5 hours for the outer region (9 exposures of 600\,s)\footnote{The reduced single pointings are available in the ESO archive and the reduced mosaic will be released soon through the ESO phase 3}.
Dithering, sky subtraction, data reduction and alignment of the two pointings were performed in the same way as for FCC\,170. \looseness-2
\begin{table*}[!h]
\centering
\caption{Uncertainties$^{\left(1\right)}$ in the calculation of stellar-kinematic and population parameters for the three galaxies FCC\,153, FCC\,177 and FCC\,170$^{\left(2\right)}$.}
\begin{adjustbox}{max width=\textwidth}

\begin{tabular}{||l|c|c|c|c|c|c|c|r||} 
\hline\hline

Name & \multirow{2}{*}{S/N} & $\delta V$ & $\delta \sigma$ & $\delta h_3$ & $\delta h_4$ & $\delta$Age & $\delta$[M/H] & $\delta$[Mg/Fe]\\
\cline{3-9}
(FCC) & & (km s$^{-1}$)& (km s$^{-1}$) & & & (Gyr) & (dex) & (dex)\\
\hline\hline
\multirow{1}{*}{153} & 60 & 4 & 5 & 0.03 & 0.04 & 2 & 0.08 & 0.05 \\
\hline
\multirow{1}{*}{177} & 60 & 5 & 6 & 0.03 & 0.04 & 2 & 0.1 & 0.05 \\
\hline
\multirow{1}{*}{170}$^{\left(3\right)}$ & 40 & 6 & 9 & 0.03 & 0.03 & 3 & 0.1 & 0.06 \\
\hline
\hline
\end{tabular}
\end{adjustbox}
\label{table:unc}
\begin{tablenotes}
\item {\footnotesize Notes. (1) The uncertainties were estimated by Monte Carlo simulations. (2) FCC\,153 and FCC\,177 are analyzed in this paper while FCC\,170 was analyzed in Paper~I and is indicated here for completeness. (3) Uncertainties from Paper~I.}
\end{tablenotes}
\end{table*}

\section{Analysis methods}
\label{sec:methods}
We used the methods described in Paper~I for FCC\,170 to extract the stellar kinematics and populations for FCC\,153 and FCC\,177.
We give here some specific details, referring to Section~4 of Paper~I for a more complete description.
Looking for a compromise between a good spatial resolution and keeping the gradients in stellar-population maps (avoiding too "noisy" maps), we chose a target signal-to-noise ratio (S/N) per spatial bin of 60 between 4750 and 5500\,\AA.
We used this value to perform the Voronoi binning using the code described in \citet{Cappellari2003}, with a minimum accepted S/N per spaxel of 1.
This resulted in 3583 bins for FCC\,153 and 3213 bins for FCC\,177.
Among them, respectively $\sim$~45\,\% and $\sim$~41\,\% had S/N\,$<60$, but only $\sim$~0.8\,\% and $\sim$~0.7\,\% had S/N\,$<50$.
We discarded 18 and 259 bins respectively for FCC\,153 and FCC\,177, because their spectra did not allow good fits or were contaminated by the light of foreground objects. \looseness-2

We fitted our spectra with the Penalized Pixel-Fitting (pPXF) code detailed in \citet{Cappellari2004} and \citet{Cappellari2017}, and MILES single-stellar-population (SSP) models from the BaSTI library, in the version with two values of [Mg/Fe]  (\citealt{Vazdekis2015}, see also Paper~I).
After checking that barely any line emission appears in the spectra of the two galaxies, we decided to perform the full-spectrum fitting by pPXF without masking any lines.
Afterwards, we calculated the residuals comparing the best fit (with no masked lines) with the original observed spectra and we confirmed that no emission lines appeared.
We fitted stellar kinematics and populations in the same run, using a multiplicative polynomial of 8th order and regularization parameter of 0.5 for both galaxies. 
We used the same approach as in Paper~I for FCC\,170 to select the level of regularization.
We verified that the features in the age-metallicity-[Mg/Fe] space do not change their position by changing the level of regularization, but they only change slightly their shape. 
\looseness-2

\section{Results}
\label{sec:results}
\subsection{Stellar kinematics} \label{sec:kin}
\begin{figure*}[!h]
\centering
\resizebox{.88\textwidth}{!}
{
\includegraphics[scale=1]{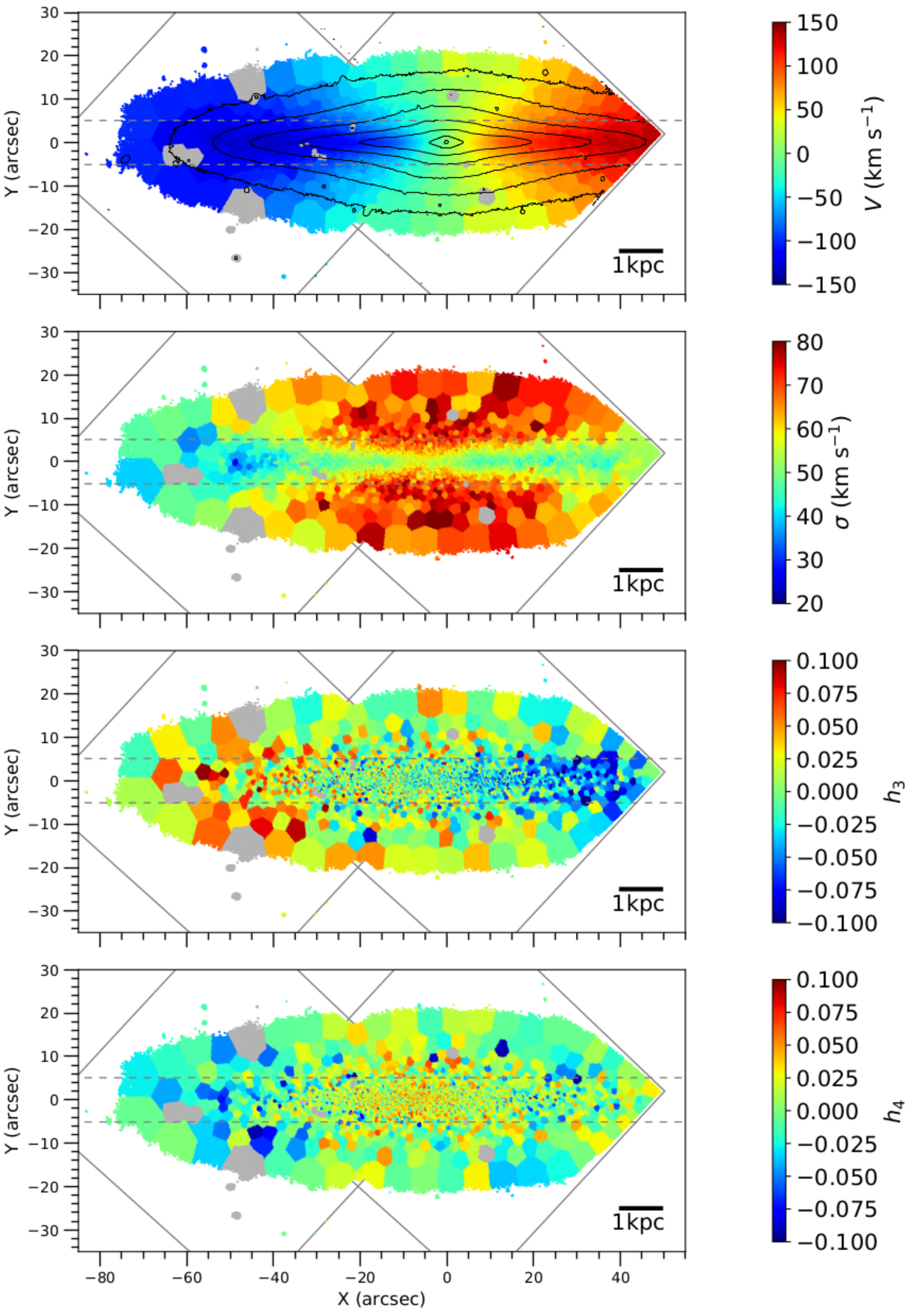}
}
\caption{Maps of the stellar kinematics of FCC\,153. From top to bottom: mean velocity $V$, velocity dispersion $\sigma$, skewness $h_3$ and kurtosis $h_4$. The discarded bins are plotted in grey, as well as the coverage of the two MUSE pointings. The physical units are indicated by the scale bar on bottom-right of each panel. The horizontal dashed lines indicate $\pm z_{\text{c1}}$. In the top panel, contours of surface brightness are plotted in black.
}
\label{chap4/fig:kin_FCC153}
\end{figure*}

The first four moments of the line-of-sight velocity distribution (LOSVD) for the two galaxies in the sample are mapped in Figs.~\ref{chap4/fig:kin_FCC153} and \ref{chap4/fig:kin_FCC177}.
Uncertainties were estimated from Monte Carlo simulations, performed as described in Appendix~A of Paper~I, and are indicated in Table~\ref{table:unc}.
These are, for each one of the two galaxies, the maximum errors obtained among 8 spatial bins in different regions of the galaxy (2 spatial bins in each one of the regions defined in \S~\ref{fig:decomp}, in general with S/N about 60). Each spectrum was perturbed 100 times with Gaussian noise and then fitted using 5 different degrees of the multiplicative polynomial around 8, and 14 different regularization parameters between $10^{-1.6}$ and 10.
In comparison to FCC\,170, whose errors are also indicated in Table~\ref{table:unc} for completeness, the uncertainties of velocity and velocity dispersion of FCC\,153 and FCC\,177, whose data were binned to a larger S/N, are slightly lower. Similar errors were found for $h_3$ and $h_4$ in the three galaxies. \looseness-2

Spider patterns appear in the velocity maps of both galaxies (top panels in Figs.~\ref{chap4/fig:kin_FCC153} and \ref{chap4/fig:kin_FCC177}). 
The maximum stellar rotation velocity is larger in FCC\,153, about 150\,\kms, than in FCC\,177, about 100\,\kms. 
Signs of cylindrical rotation are not clear in FCC\,153, in agreement with the fact that its bulge is not a prominent bar seen edge-on. 
In
FCC\,177, this cylindrical rotation is visible in a box of side of about 20\,arcsec parallel to the midplane, corresponding to the boxy bulge.
Neither of the two galaxies shows a central peak in velocity dispersion, as seen in the second panels from the top in both Figs.~\ref{chap4/fig:kin_FCC153} and \ref{chap4/fig:kin_FCC177}. This confirms that we do not have any massive central concentration of stars (e.g., any prominent bulge).
The highest $\sigma$ values are measured in the thick disk region (dominating above $z_{\text{c1}}$ from Table~\ref{chap4/table:gal_prop}).
In FCC\,153, the dynamically cold thin disk is well defined with a sharp transition to the dynamically hotter thick disk. It displays a clear flare which starts not far  (at about 5\,arcsec) from the galaxy center, implying a radial gradient of velocity dispersion in the outer region of the (geometrically-defined) thick disk. The thin disk also shows a negative $\sigma$ gradient up to a certain galactocentric distance (about 50-60\,arcsec), while a small increase is suggested farther in radius.
In contrast, the dynamically cold region is much thicker in FCC\,177, where a $\sigma$ vertical gradient is still present above $z_{\text{c1}}$. A flaring is not clear, probably also because our data does not go far enough from the midplane in the external part, where this flaring seems to occur as suggested by the most external spatial bins in our maps. \looseness-2
\begin{figure*}[!h]
\centering
\resizebox{.75\textwidth}{!}
{
\includegraphics[scale=1]{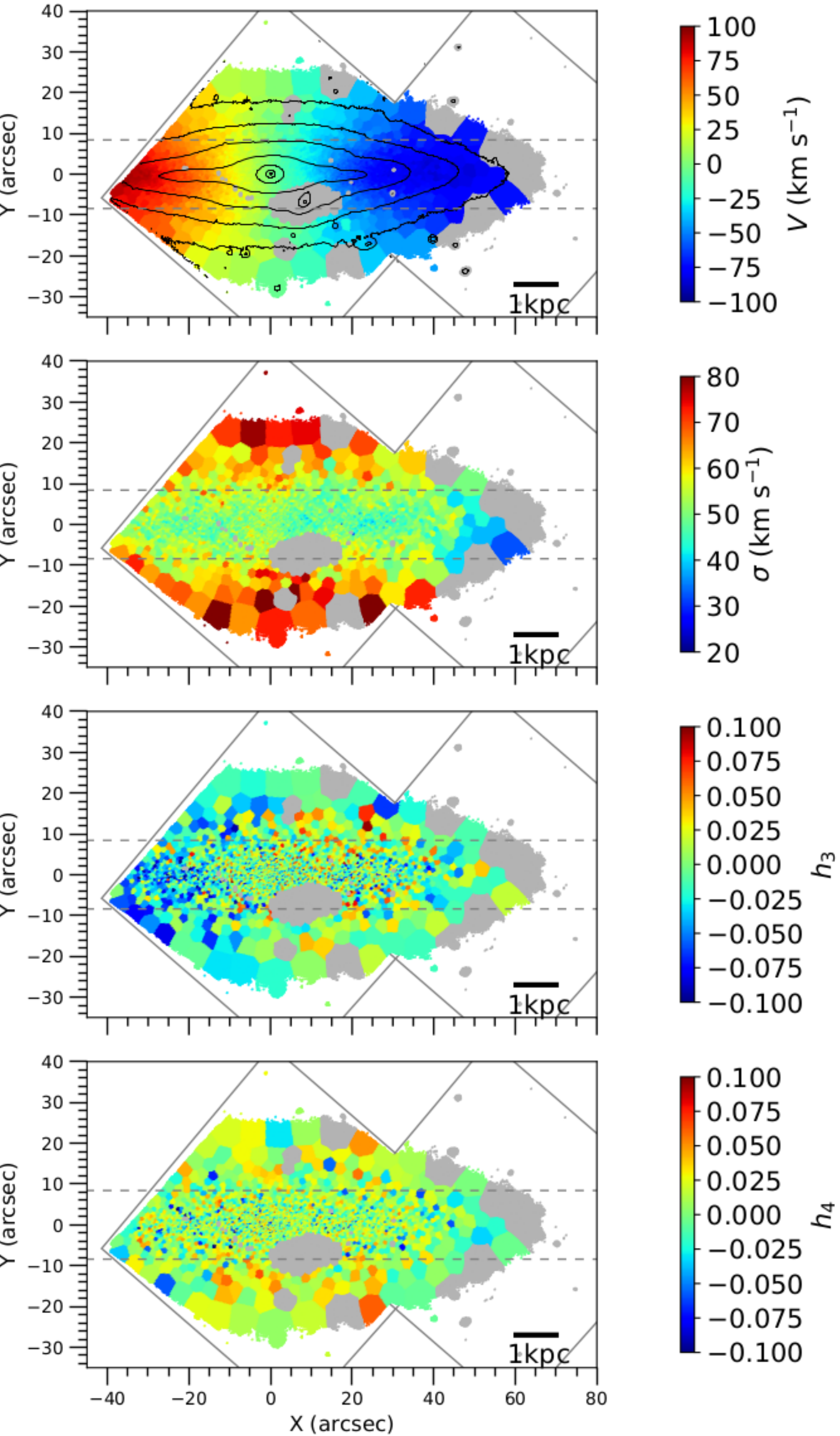}
}
\caption{As Fig.~\ref{chap4/fig:kin_FCC153}, now for FCC\,177.
}
\label{chap4/fig:kin_FCC177}
\end{figure*}

The third-from-top panel of Fig.~\ref{chap4/fig:kin_FCC153} show an anticorrelation of $h_3$ with $V$, associated to a disk-like rotation in FCC\,153 \citep[e.g.,][]{Krajnovic2008,Guerou2016}. $h_4$ (bottom panel of Fig.~\ref{chap4/fig:kin_FCC153}) assumes in the central region of FCC\,153 positive values, associated to a LOSVD with
a narrower symmetric profile with respect to a pure Gaussian \citep{vanderMarel1993,Gerhard1993}. Negative values are displayed in the more external radii near the midplane.
In FCC\,177, no structures are clearly seen in the $h_3$ and $h_4$ maps (see two bottom panels of Fig.~\ref{chap4/fig:kin_FCC177}) and even the disk $h_3-V$ anticorrelation is barely visible. \looseness-2

\begin{figure*}[!h]
\begin{minipage}[H!]{1\textwidth}
\centering
\resizebox{1\textwidth}{!}
{
\includegraphics[scale=1]{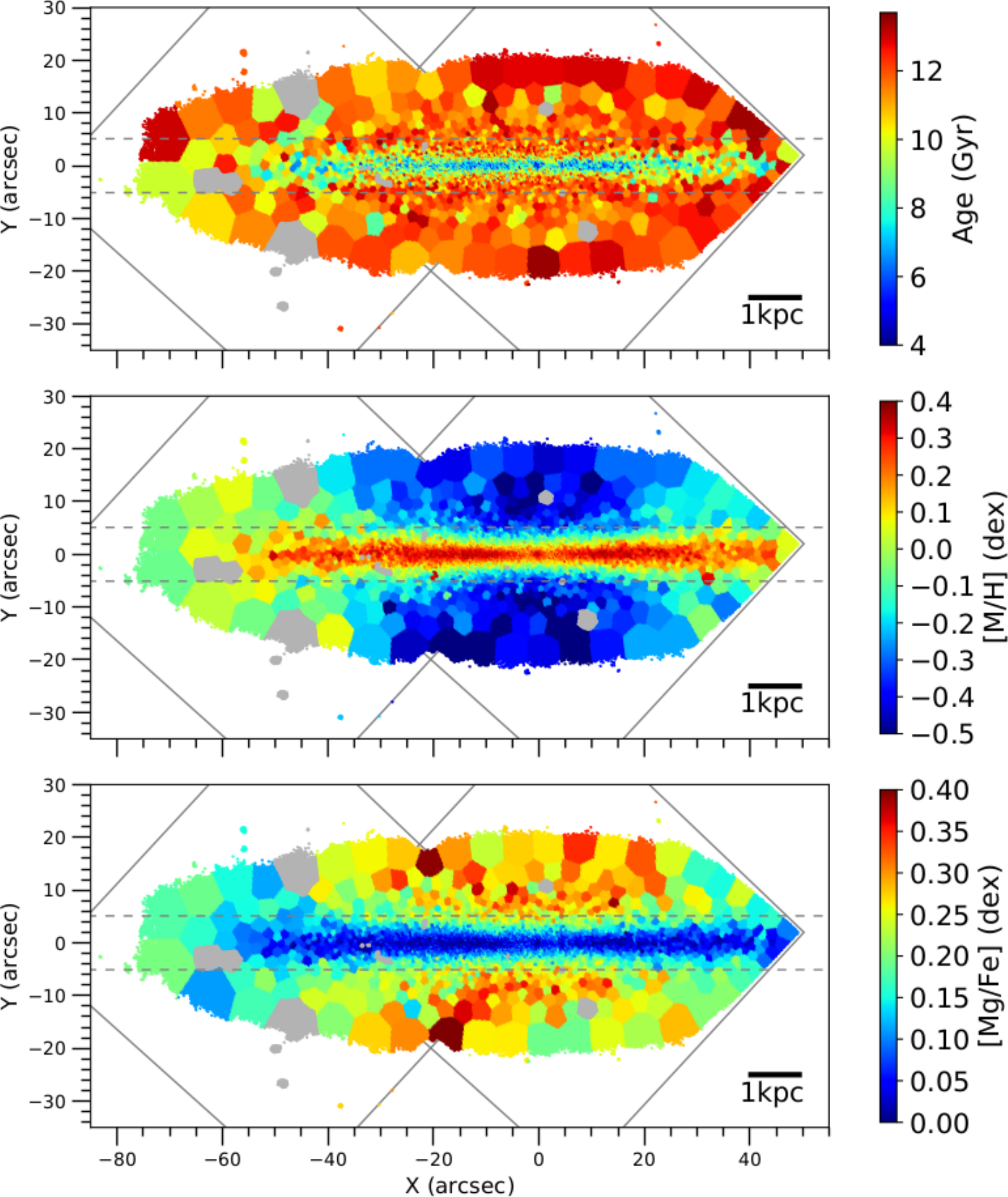}
}
\caption{
Stellar-population maps for FCC\,153.
From top to bottom: mean age, total metallicity [M/H], [Mg/Fe] abundance. The discarded bins are plotted in grey, as well as the coverage of the two MUSE pointings. The physical units are indicated by the scale bar on bottom-right of each panel. The horizontal dashed lines indicate $\pm z_{\text{c1}}$.
}
\label{chap4/fig:pop_FCC153}
\end{minipage}
\end{figure*}
\begin{figure*}[!h]
\centering
\resizebox{.88\textwidth}{!}
{
\includegraphics[scale=1]{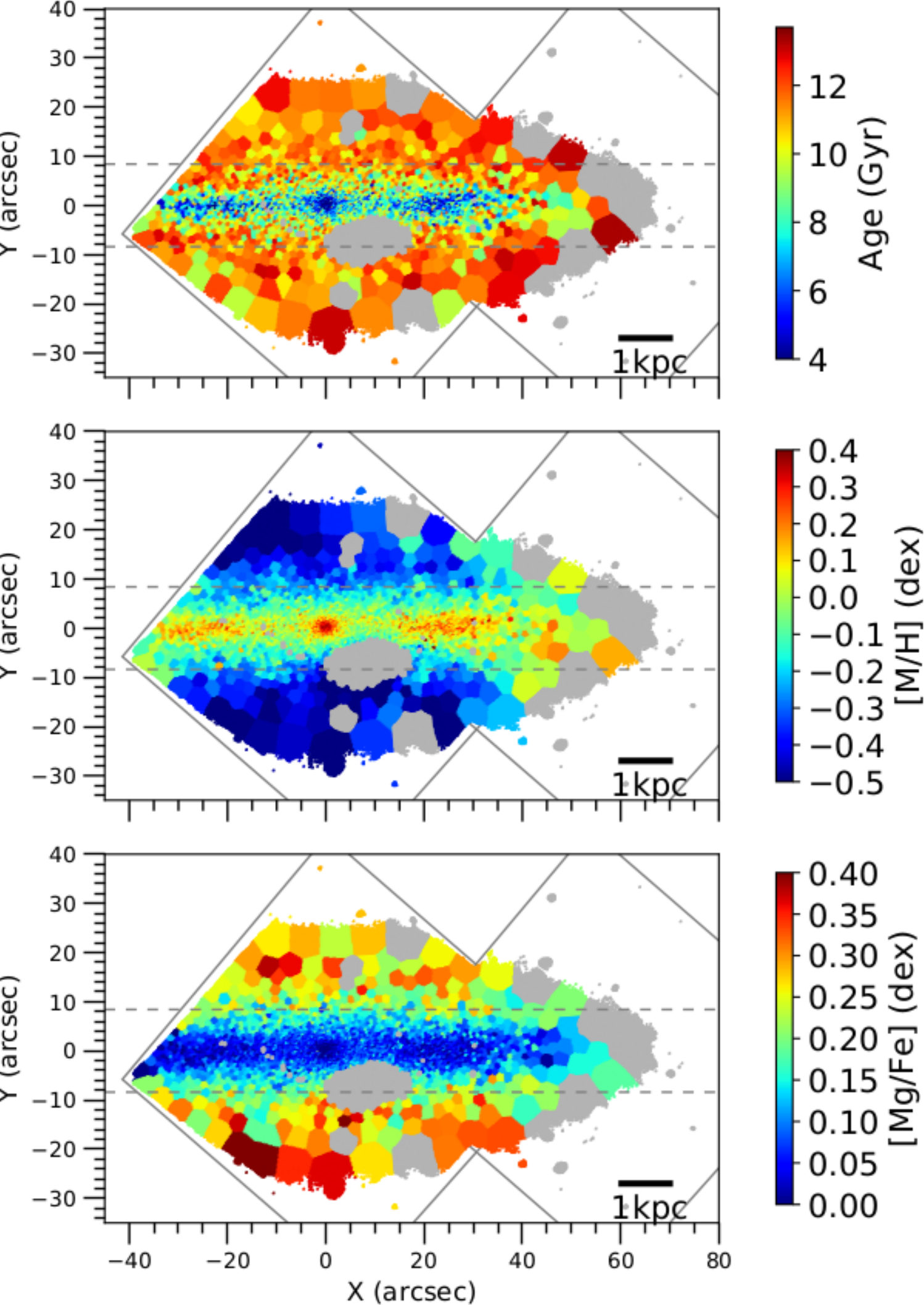}
}
\caption{
As Fig.~\ref{chap4/fig:pop_FCC153}, now for FCC\,177.
}
\label{chap4/fig:pop_FCC177}
\end{figure*}
\begin{table*}[!h]
\centering
\caption{Definition of the structural components of FCC\,153 and FCC\,177: nucleus, thin disk, inner and outer thick disk.}
\begin{adjustbox}{max width=\textwidth}

\begin{tabular}{||l|c|c|c|c|c|c|c|r||} 
\hline\hline
\centering 
\multirow{3}{*}{FCC} & \multicolumn{2}{c}{\multirow{2}{*}{Nucleus}} & \multicolumn{2}{|c}{\multirow{2}{*}{Thin disk}} & \multicolumn{4}{|c||}{Thick disk}\\
\cline{6-9}
& \multicolumn{2}{c}{} &\multicolumn{2}{|c}{} &\multicolumn{2}{|c|}{Inner} & \multicolumn{2}{|c||}{Outer}\\
\cline{2-9}
& $\abs{x}$ (\arcsec) & $\abs{y}$ (\arcsec)& $\abs{x}$ (\arcsec)& $\abs{y}$ (\arcsec)& $\abs{x}$ (\arcsec)& $\abs{y}$ (\arcsec)&$\abs{x}$ (\arcsec)& $\abs{y}$ (\arcsec)\\
\hline\hline
153 & $< 0.2$ & $< 0.2$ & $>6$ & $<2$ & $<20$ & $>6$  & $>20$ & $>6$ \\
\hline
177 & $<0.4$ & $<0.4$  & $>13$ & $<4$ & $<25$ & $>10$ & $>25$ & $>10$ \\
\hline
\hline
\end{tabular}
\end{adjustbox}
\label{table:str_dec}
\end{table*}

\subsection{Mass-weighted stellar populations} \label{sec:pop}
The full-spectrum-fitting method used in this work allows us to extract age, total metallicity ([M/H]) and [Mg/Fe] abundance.
It gives as output the weights assigned to the individual SSP models, from which we recover the weighted average stellar-population properties.
This method allows us to obtain average values of [Mg/Fe] in the full range between 0.0 and 0.4\,dex, in spite of having SSP models with only these two values.
Since [Mg/Fe] is an indicator of $\alpha$-element abundances, for simplicity we will refer to its enhancement also as ''$\alpha$-enhancement'', often interpreted as due to a shorter star-formation timescale.
We show the maps of the mean stellar-population properties of FCC\,153 and FCC\,177 in Figs.~\ref{chap4/fig:pop_FCC153} and \ref{chap4/fig:pop_FCC177}.
Estimates of uncertainties from Monte Carlo simulations (see \S\ref{sec:kin}) are indicated in Table~\ref{table:unc}. Similar uncertainties were found for the stellar-population parameters of FCC\,153 and FCC\,177.
We found uncertainties similar to FCC\,170 for [M/H] and [Mg/Fe], but lower for the mean stellar age. \looseness-2

The different structural components of the two galaxies are remarkably clear in Figs.~\ref{chap4/fig:pop_FCC153} and \ref{chap4/fig:pop_FCC177}.
In both top panels, there is a clear age distinction between the thin and the thick disk, the latter being very old (about 12--13\,Gyr).
Young stars are present close to the midplane, in a much narrower band in FCC\,153 than in FCC\,177.
The thin-disk youngest stars are concentrated in a certain radial range (mainly within about 20 and 30-40\,arcsec respectively in FCC\,153 and FCC\,177) and a positive age gradient is visible in both galaxies along the midplane.
Younger spatial bins in the thick-disk outer region of FCC\,153 are probably connected to the thin-disk flaring detected in the velocity dispersion map.
The youngest stars are distributed in the center of both galaxies, in a spherical/elliptical shape much larger than the NSC detected by \citet{Turner2012} (see also Table~\ref{chap4/table:gal_prop}).
This is very likely due to the larger PSF (Point Spread Function) in our MUSE data than in data from the Advanced Camera for Surveys mounted at the \textit{Hubble Space Telescope}, used by \citet{Turner2012}.
We have measured ages down to $\sim$\,4\,Gyr in FCC\,153, and down to $\sim$\,2\,Gyr in FCC\,177 in their nuclear regions (hereafter "nuclei"), which include not only the pure stellar populations of the NSC but all components observed in the line of
sight. On the other hand, their thin disks appear approximately coeval with one another ($\sim 4-5$\,Gyr old in their youngest regions). \looseness-2

In both galaxies there is a clear correspondence between age and metallicity (top and middle panels in Figs.~\ref{chap4/fig:pop_FCC153} and \ref{chap4/fig:pop_FCC177}), with the younger stars being more metal rich and vice versa.
The nuclei are clearly visible in the metallicity maps, where they assume the highest (supersolar) values.
The thin disk is very thin and well defined in FCC\,153's metallicity map.
It is thicker and fuzzier in FCC\,177. This difference between the two galaxies is consistent with their velocity-dispersion maps.
Radial gradients are present in disks of both galaxies. A negative metallicity gradient is seen in thin disks, while the thick-disk subsolar metallicities increase towards the outskirts.
The variation of this gradient with height in FCC\,153 suggests that it is strongly related to the thin-disk flare.
We found similar values of the total metallicity in the thick disks of the two galaxies, while FCC\,153's thin disk appears more metal rich than the one in FCC\,177. \looseness-2

The maps of [Mg/Fe] (bottom panels in Figs.~\ref{chap4/fig:pop_FCC153} and \ref{chap4/fig:pop_FCC177}) show distinctions, between the thin and the thick disks, similarly to age and metallicity maps.
Midplane stars display abundances around solar (including the thin disk and the nucleus), while thick-disk stars are quite $\alpha$-enhanced.
Also, the behavior of the [Mg/Fe] radial gradients are inverted with respect to metallicity: positive in the outer thin disk, but negative in the thick disk (one more time, very clear in FCC\,153 and not so clear in FCC\,177). \looseness-2

\subsection{Structural components}\label{sec:decomp}
\begin{figure}[!h]
\scalebox{0.48}{
\includegraphics[scale=0.89]{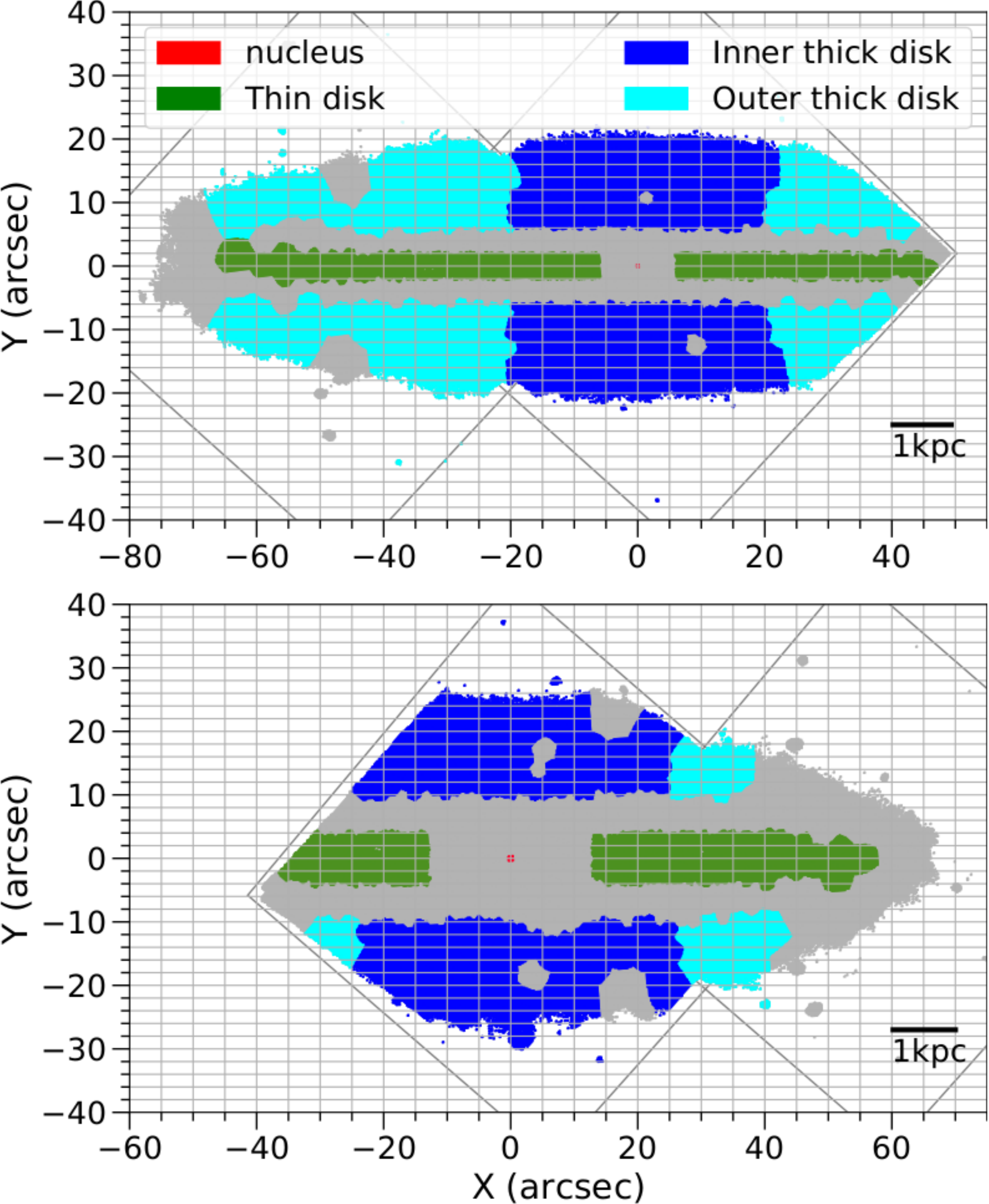}
}
\caption{Definition of the structures in FCC\,153 (top panel) and FCC\,177 (bottom panel). The spatial bins are color coded for each one of the components. The thick disk is divided into the inner and outer regions. The spatial bins not taken into account for the study of the individual components (bad bins or bins in the transition where more than one components are important) and the coverage of the two MUSE pointings are plotted in grey. The scale bar on bottom-right of each panel indicates the correspondence with physical units.}
\label{fig:decomp}
\end{figure}

For the rest of the paper, we define the structures of FCC\,153 and FCC\,177 more accurately.
We based our definition of the components on the previous photometric analysis by \citet{Comeron2018} and \citet{Turner2012} (see also Section~\ref{sec:sample}), using a similar approach as for FCC\,170 (Paper~I) and being somewhat more restrictive for the thick disk than the limits in Table~\ref{chap4/table:gal_prop}.
We divided the spatial bins according to which of the three main components dominates: nucleus, thin disk, and thick disk.
We excluded a region of transition between the thin and thick disks, based on our population maps.
Given that we could not distinguish the stellar populations of the bulges in our maps, we considered that it was not possible to analyze their properties with the line-of-sight integrated data. 
However, from the kinematic maps, we were able to define a central region which could be influenced by the bulge (e.g. where the mean-velocity spider pattern is not clear or the velocity dispersion pattern is distorted). We discarded this region from the analysis of the thin and thick disks in both galaxies.
We also divided the thick-disk bins into inner and outer regions, given that they have different kinematic and chemical properties.
We selected these two regions according to the maps of $\sigma$, [M/H] and [Mg/Fe], defining the inner thick disk as the region where properties were approximately (spatially) constant.
In this way, the inner thick disk would be a region of the thick disk not affected by the thin-disk flaring. In contrast, we expect the outer thick disk to be contaminated by thin-disk-like populations.
The limits between the components are indicated in Table~\ref{table:str_dec} and mapped in Fig.~\ref{fig:decomp}. \looseness-2

\subsection{Star formation history} \label{sec:SFH}
\begin{figure*}[!h]
\centering
\resizebox{0.65\textwidth}{!}
{
\includegraphics[scale=0.49]{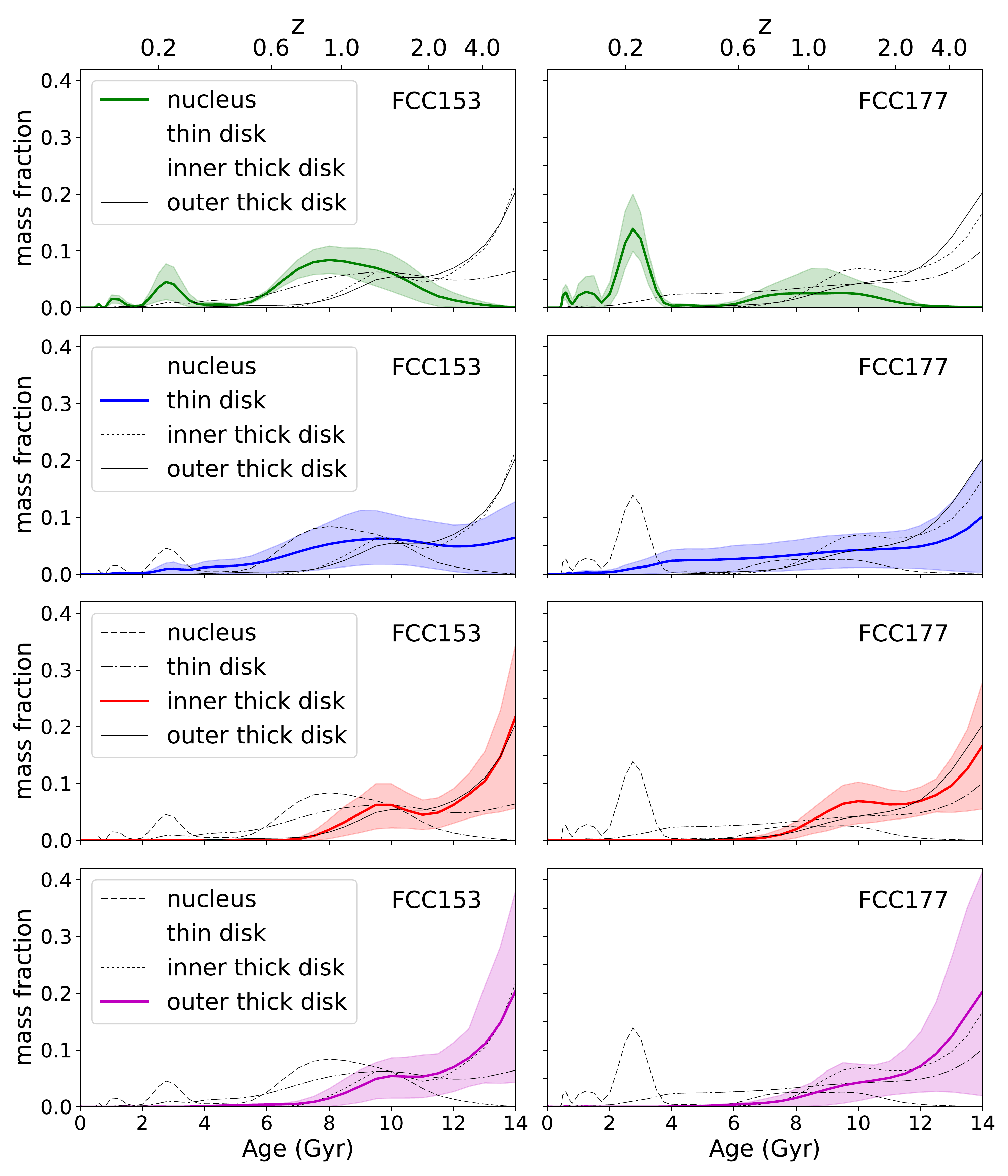}
}
\caption{
SFH of the structural components of FCC\,153 (left column) and FCC\,177 (right column) defined in Fig.~\ref{fig:decomp}.
The average mass fraction is displayed on the vertical axis. 
The average was calculated on the different spatial bins and was weighted with the mass in each one of them.
This mass fraction is normalized to the total mass of the component.
In each panel, only one component is represented in a specific color (\textit{green} for the nucleus, \textit{blue} for the thin disk, and \textit{red} and \textit{purple} respectively for the inner and the outer thick disk). For the colored component, 1$\sigma$ uncertainties, calculated as 16\,\% and 84\,\% percentiles of the spatial-bin distribution, are represented as shades of the same color. The SFH of the rest of components are plotted in black, with different line styles, in each panel.
As reference, the redshift scale on top of each column indicates the epoch when stars of a given age were born. The correspondence between redshift and age was approximated with Eq.~2 in \citet{Carmeli2006}.
}
\label{chap4/fig:sfh}
\end{figure*}

\begin{figure*}[!h]
\centering
\resizebox{0.65\textwidth}{!}
{
\includegraphics[scale=0.49]{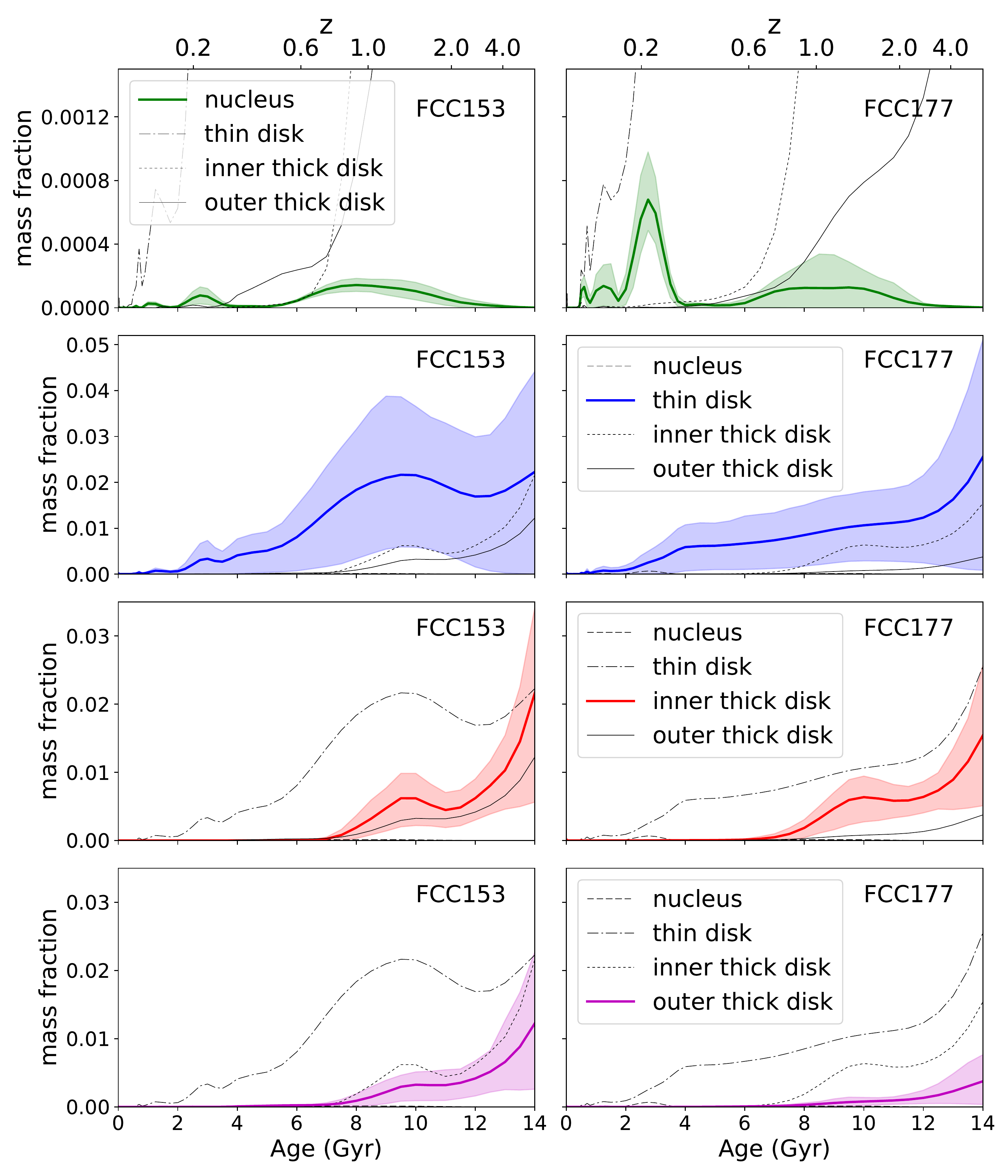}
}
\caption{
As in Fig.~\ref{chap4/fig:sfh} but now the mass fraction is normalized to the total mass of the galaxy.
}
\label{chap4/fig:sfh_normtot}
\end{figure*}

The geometrical definition of the structural components allows us to analyze individually the star formation history (SFH) of the nucleus, the thin, and the thick disk of the two galaxies.
We calculated the SFHs of FCC\,153 and FCC\,177 using the same method used in Paper~I.
The mass fractions assigned by pPXF to coeval models were summed to obtain the SFH of each spatial bin, understood as the mass fraction of each spatial bin as function of age.
For each individual structural component, we averaged the SFHs of all the spatial bins, weighting by their mass. 
From the mass in every spatial bin, we estimated the total mass of the two galaxies within the regions covered by our MUSE pointings (discarded bins excluded). This lower limit for the total galaxy mass is $\sim 2.9\times 10^{10}$\,M$_{\odot}$ for FCC\,153 and $\sim 1.1\times 10^{10}$\,M$_{\odot}$ for FCC\,177.
These values are consistent with the ones in Table~\ref{chap4/table:gal_prop}, although these are lower (especially for FCC\,153). 
Our values and the ones from \citet{Iodice2019} were calculated using completely different methods. Approximations were used in both cases, providing an additional source of uncertainty.
We show the SFH of the different components of the two galaxies in Figs.~\ref{chap4/fig:sfh} and \ref{chap4/fig:sfh_normtot}.
In the former, the mass fraction is normalized to the total mass of the component, so it is easier to compare the different populations in the same component. In the latter, the mass fraction is normalized to the total mass of the galaxy and we can compare between the different components. \looseness-2

The nuclei exhibit "bursty" SFHs, in the sense that they formed during different intense episodes (green curves in the top panels of Figs.~\ref{chap4/fig:sfh} and \ref{chap4/fig:sfh_normtot}). 
The nuclear region appears younger in FCC\,177 than in FCC\,153. 
The nucleus of FCC\,177 formed relatively few stars at a stable rate between 6 and 12\,Gyr ago, but had a later intense burst between 2 and 4\,Gyr.
FCC\,153 formed more stars in its nucleus in an early epoch, although it had some other recent peaks in the star formation.
The thin disks present a continuous SFH, as shown in the second from top panels of Figs.~\ref{chap4/fig:sfh} and \ref{chap4/fig:sfh_normtot}.
In both galaxies, either thin disks started to form stars very early or line-of-sight effects make the thick disk stars contribute for the oldest ages.
Both thin disks continued to grow their stellar mass up to recent times (about 2\,Gyr). 
The thick disks have very similar SFH in the two galaxies (two bottom panels in Figs.~\ref{chap4/fig:sfh} and \ref{chap4/fig:sfh_normtot}).
They formed early, with a main component $12-14$\,Gyr old. However, an additional population appeared later, with 10-Gyr-old stars, indicated by a peak in the SFH.
This second younger component is more prominent in the inner thick disk than in the outer thick disk, where we probably have a mixing with other populations due to the thin-disk flaring. This makes the SFH appear slightly smoother in the outer thick disk. \looseness-2

\subsection{Chemical abundances} \label{sec:chem}
\begin{figure*}[!h]
\centering
\resizebox{.7\textwidth}{!}
{
\includegraphics[scale=0.35]{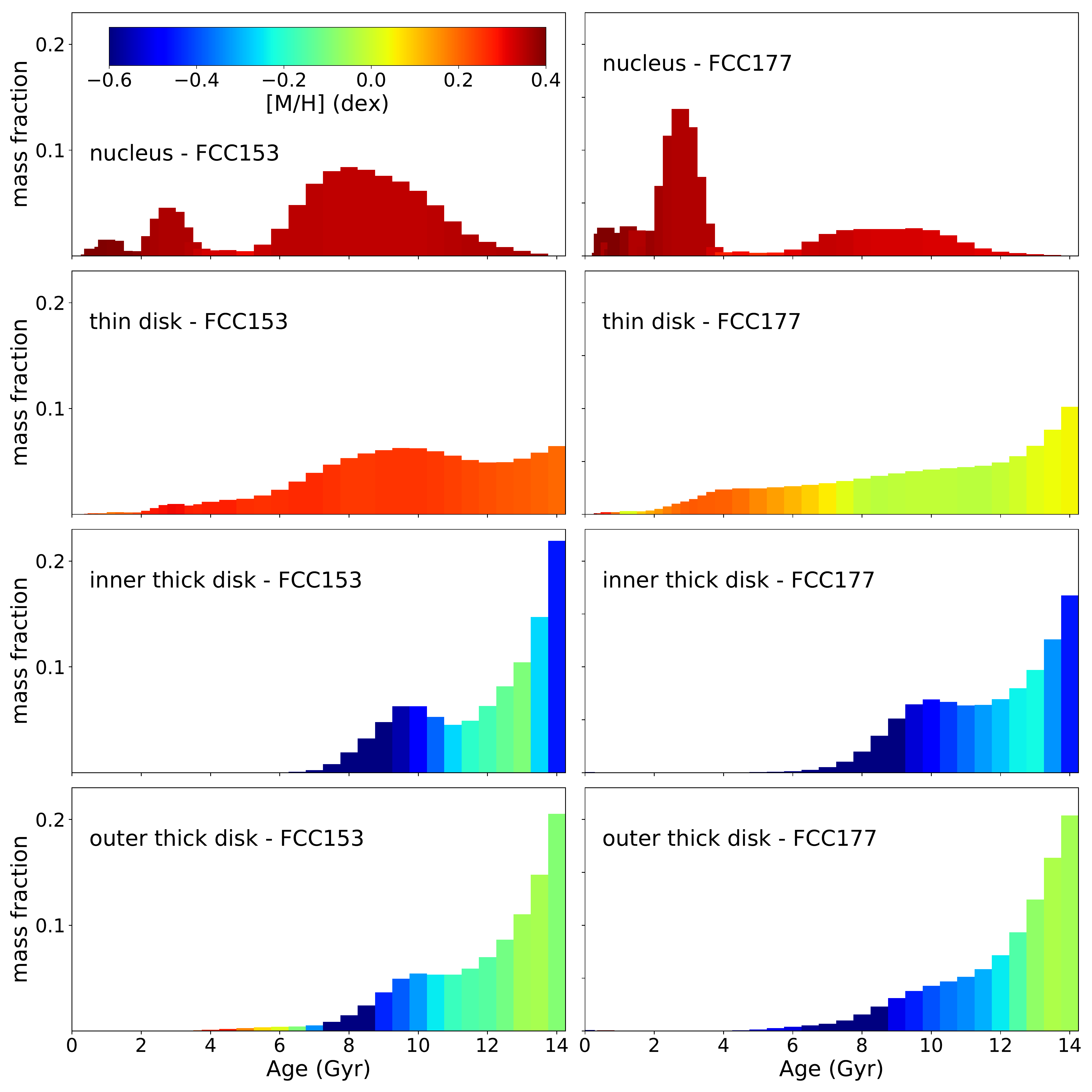}
}
\caption{SFH of the different structural components of FCC\,153 (left panels) and FCC\,177 (right panels).
From top to bottom: the nucleus, the thin disk, the inner and the outer thick disk.
The histogram age bins are color coded according to the weighted average total metallicity ([M/H]) of the specific age bin.
The average mass fraction is displayed on the vertical axis, weighted by the mass in each bin  and normalized to the mass of the component.
}
\label{chap4/fig:agemet_hist}
\end{figure*}

\begin{figure*}[!h]
\centering
\resizebox{.7\textwidth}{!}
{
\includegraphics[scale=0.35]{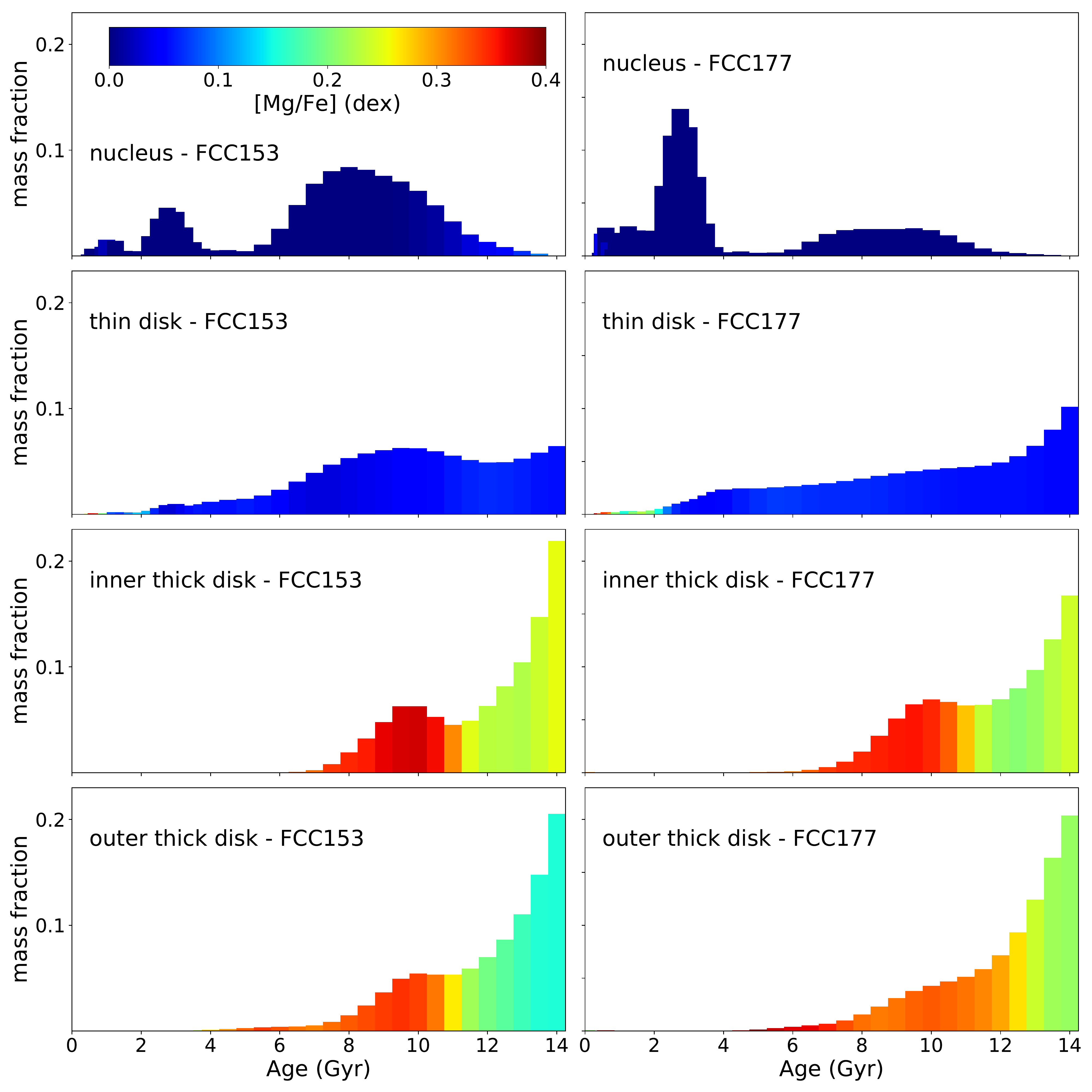}
}
\caption{As in Fig.~\ref{chap4/fig:agemet_hist}, but now the histogram age bins are color coded according to the weighted average [Mg/Fe] of the specific age bin.
}
\label{chap4/fig:agealpha_hist}
\end{figure*}

We performed a chemical analysis of the different populations by color coding the age bins of the SFH histograms according to their chemical properties.
In Fig.~\ref{chap4/fig:agemet_hist} and \ref{chap4/fig:agealpha_hist}, we show the SFH of FCC\,153 (left panels) and FCC\,177 (right panels), indicating the mean [M/H] and [Mg/Fe], respectively.
We used the same normalization as in Fig.~\ref{chap4/fig:sfh}, with respect to the mass of the structural component.
Nuclei, the most metal-rich components, display slightly higher metallicities for the youngest stars than for the rest of ages (top panels in Fig.~\ref{chap4/fig:agemet_hist}). They display also the lowest [Mg/Fe] abundance of all galactic components, around solar values (top panels in Fig.~\ref{chap4/fig:agealpha_hist}). These nuclear regions do not show a strong time evolution in the mean [M/H] and [Mg/Fe].
Thin disks present different initial (average) metallicities for the two galaxies (second panels from top, in Fig.~\ref{chap4/fig:agemet_hist}). FCC\,153 started already with a metal-rich thin disk (at older ages), whose interstellar medium seems not to have evolved much since then.
On the contrary, FCC\,177's thin disk exhibits a clear positive metallicity evolution with time, from about solar values to clearly supersolar.
This indicates a gradual chemical enrichment.
In relation to their [Mg/Fe], both thin disks display uniform slightly supersolar values over time (see second panels from top, in Fig.~\ref{chap4/fig:agealpha_hist}). The gradual chemical evolution of FCC\,177's thin disk is not visible in the [Mg/Fe] abundance. \looseness-2

Inner thick disks are the most metal-poor and [Mg/Fe]-enhanced components in both galaxies (third from top panels in Figs.~\ref{chap4/fig:agemet_hist} and \ref{chap4/fig:agealpha_hist}), with a big difference in values if compared to the respective nuclei and thin disks (up to 1~dex in metallicity and 0.4~dex in [Mg/Fe]).
Similarly to what was found for FCC\,170, the thick disks in FCC\,153 and FCC\,177 display a dominant very old population and a smaller contribution of younger stars.
The oldest populations were born with relatively low metallicity, but the gas was slightly enriched over time, increasing the mean metallicity of later stars slightly.
While this was occurring, the 
younger populations (corresponding to the secondary peak in the thick-disk SFH) were being formed.
These younger components
 have similar properties in the two analyzed galaxies. 
They are approximately 10\,Gyr old (probably slightly younger in FCC\,153) and they have similar metallicities and [Mg/Fe] abundances.
These populations brought a new chemistry to the galaxies: the lowest metallicity (the darkest blue in Fig.~\ref{chap4/fig:agemet_hist}) and the highest [Mg/Fe] (red color in Fig.~\ref{chap4/fig:agealpha_hist}).
The outer regions of the thick disks (bottom panels in Figs.~\ref{chap4/fig:agemet_hist} and \ref{chap4/fig:agealpha_hist}), which include the spatial bins located in the thin-disk flare, appear in general slightly less metal-poor and $\alpha$-enhanced than the inner regions.
Probably related to the flare, the difference in the [Mg/Fe] abundance between the inner and the outer thick disk is more pronounced  in FCC\,153 than in FCC\,177. \looseness-2

\begin{figure*}[!h]
\centering
\resizebox{1\textwidth}{!}
{
\includegraphics[scale=0.9,width=\textwidth]{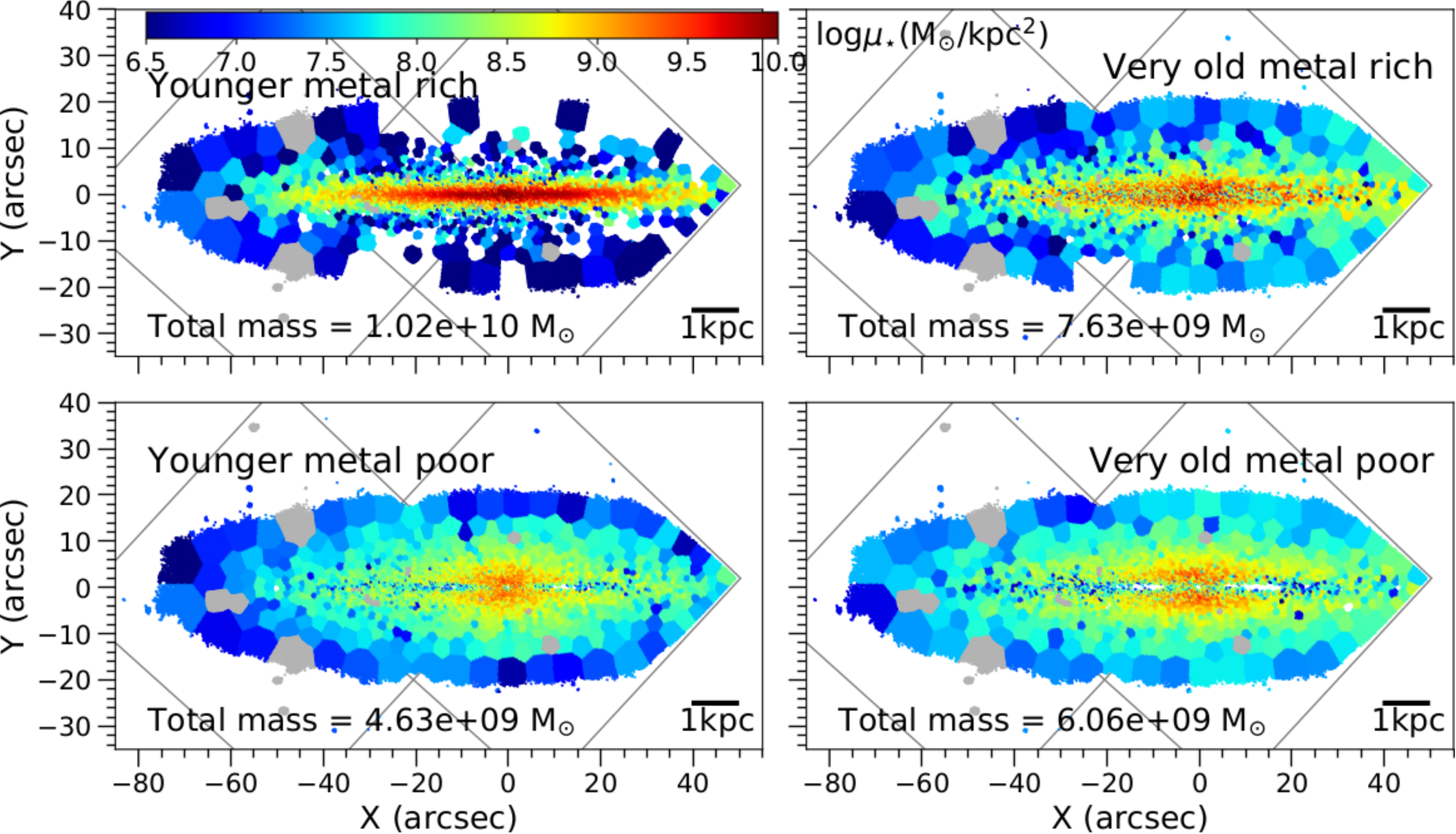}
}
\caption{Spatial distribution of stars of a given range of age and metallicity, in FCC\,153.
\textit{Top}: most metal-rich populations ([M/H]$\ge 0.06$\,dex).
\textit{Bottom}: most metal-poor populations ([M/H]$\le -0.25$).
\textit{Left}: slightly younger populations (ages $\le$\,11.0\,Gyr).
\textit{Right}: oldest populations (ages $\ge$\,11.5\,Gyr).
The color scale shows the mass density corresponding to the populations in the specific age-metallicity
bin.
The total mass in the age-metallicity bin is indicated on bottom-left of each map.
The coverage of the two MUSE pointings is plotted in grey as well as the discarded bins. A scale bar on bottom-right
of each map indicates the correspondence with physical units.
}
\label{fig:FCC153_agemet_bins}
\end{figure*}

\begin{figure*}[!h]
\centering
\resizebox{1\textwidth}{!}
{
\includegraphics[scale=0.9,width=\textwidth]{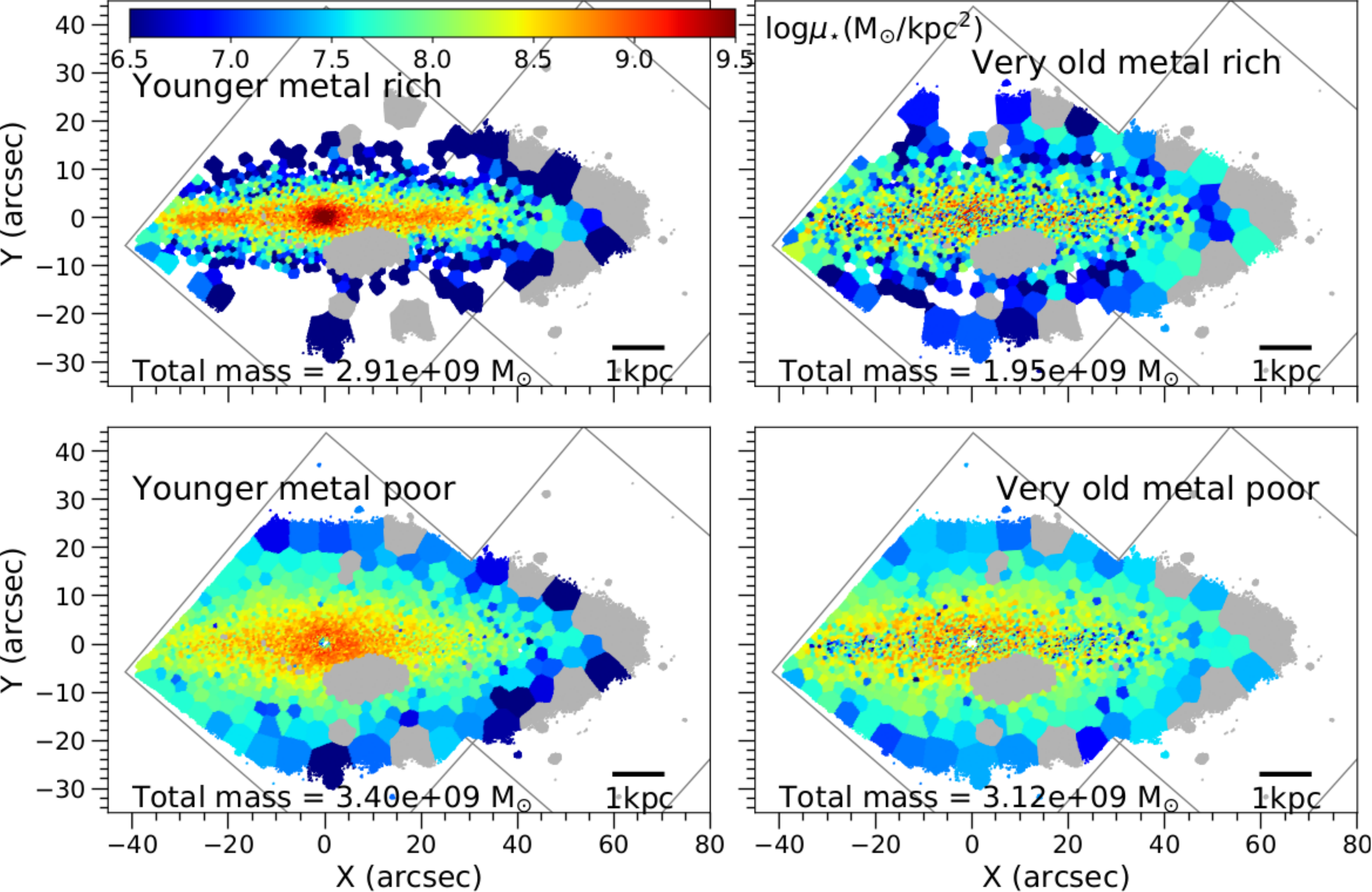}
}
\caption{As Fig.~\ref{fig:FCC153_agemet_bins}, now for FCC\,177.
}
\label{fig:FCC177_agemet_bins}
\end{figure*}

We can look at the spatial distribution of the different populations mapping the stars in different age and metallicity ranges. 
We used this approach to produce Figs.~\ref{fig:FCC153_agemet_bins}, \ref{fig:FCC177_agemet_bins}, \ref{appA/fig:age_bins_FCC153}, \ref{appA/fig:age_bins_FCC177}, \ref{appA/fig:met_bins_FCC153} and \ref{appA/fig:met_bins_FCC177}.
In FCC\,153, the metal-rich stars are concentrated in the thin disk (supersolar metallicity, top panels of Fig.~\ref{fig:FCC153_agemet_bins}). In particular, the relatively young populations (younger than 11\,Gyr, top-left panel) are concentrated right in the midplane and absent in numerous bins of the inner thick disk.
The populations with subsolar metallicity (bottom panels of Fig.~\ref{fig:FCC153_agemet_bins}) are distributed all over the galaxy. However, they show a higher density close to the galaxy center but lower density right in the midplane, dominated by the metal-rich stars (this effect is more visible in the bottom-right panel).
Similar results are obtained for FCC\,177 (Fig.~\ref{fig:FCC177_agemet_bins}). Here, the younger metal-rich stars are located mainly in the nucleus, and to a minor extent in the thin disk (top-left panel). The metal-rich but older stars have a more uniform density in a much thicker region aligned with the major axis (top-right panel), than in FCC\,153. \looseness-2

\begin{figure*}[!h]
\centering
\resizebox{0.8\textwidth}{!}
{
\includegraphics[scale=1,width=\textwidth]{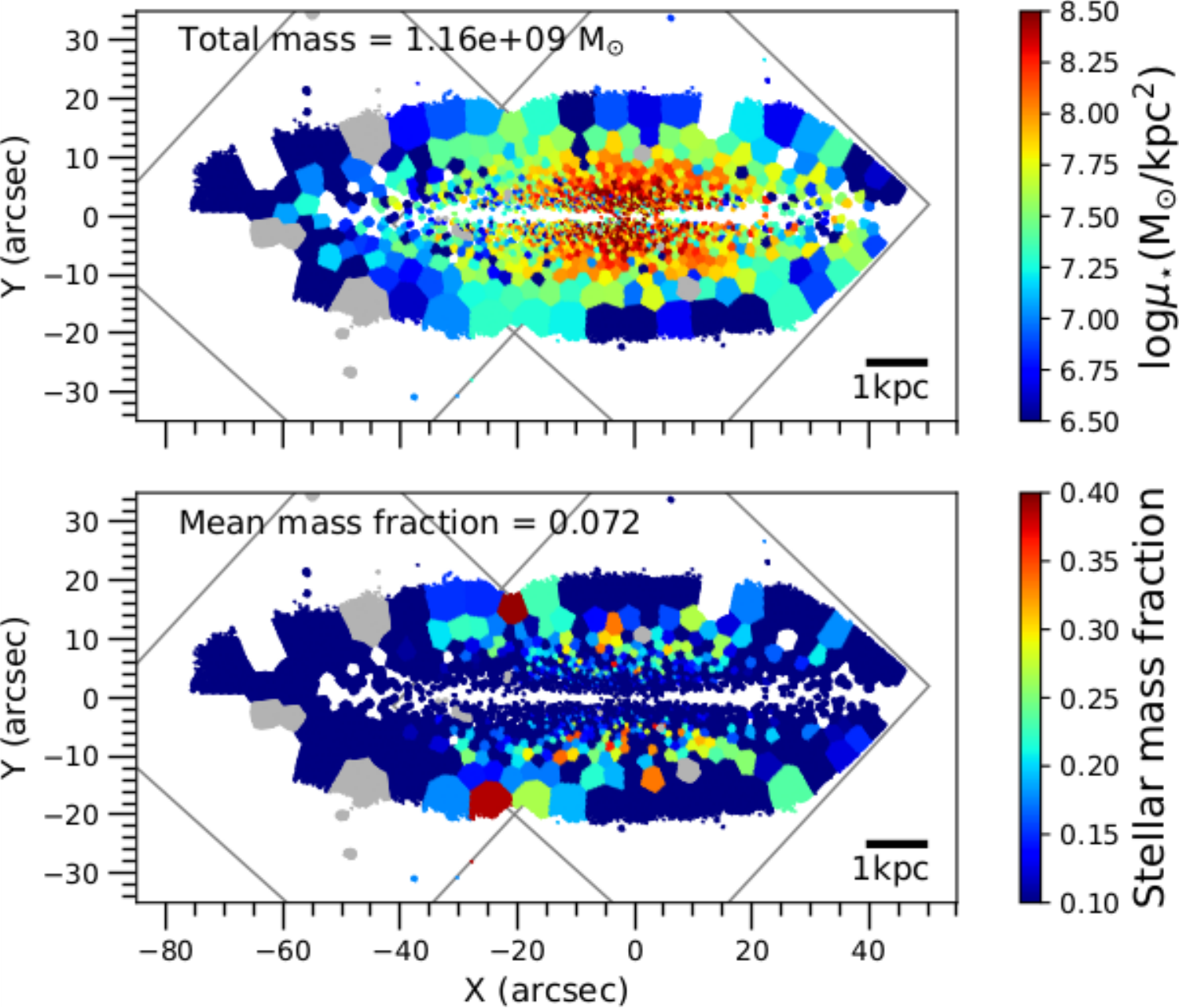}
}
\caption{Distribution in FCC\,153 of the stellar population with ages between 6 and 11\,Gyr, metallicities between -0.6 and -0.35\,dex and [Mg/Fe]\,$\sim 0.4$\,dex.
The color scale shows the stellar mass density in the top panel and the stellar mass fraction in the bottom panel.
The total mass or the mean mass fraction in this specific population are indicated on top-left of respectively the top and the bottom maps.
In each panel, the coverage of the two MUSE pointings and the discarded bins are plotted in grey. A scale bar on bottom-right indicates the correspondence with physical units.
}
\label{fig:FCC153_accreted}
\end{figure*}

\begin{figure*}[!h]
\centering
\resizebox{0.8\textwidth}{!}
{
\includegraphics[scale=1,width=\textwidth]{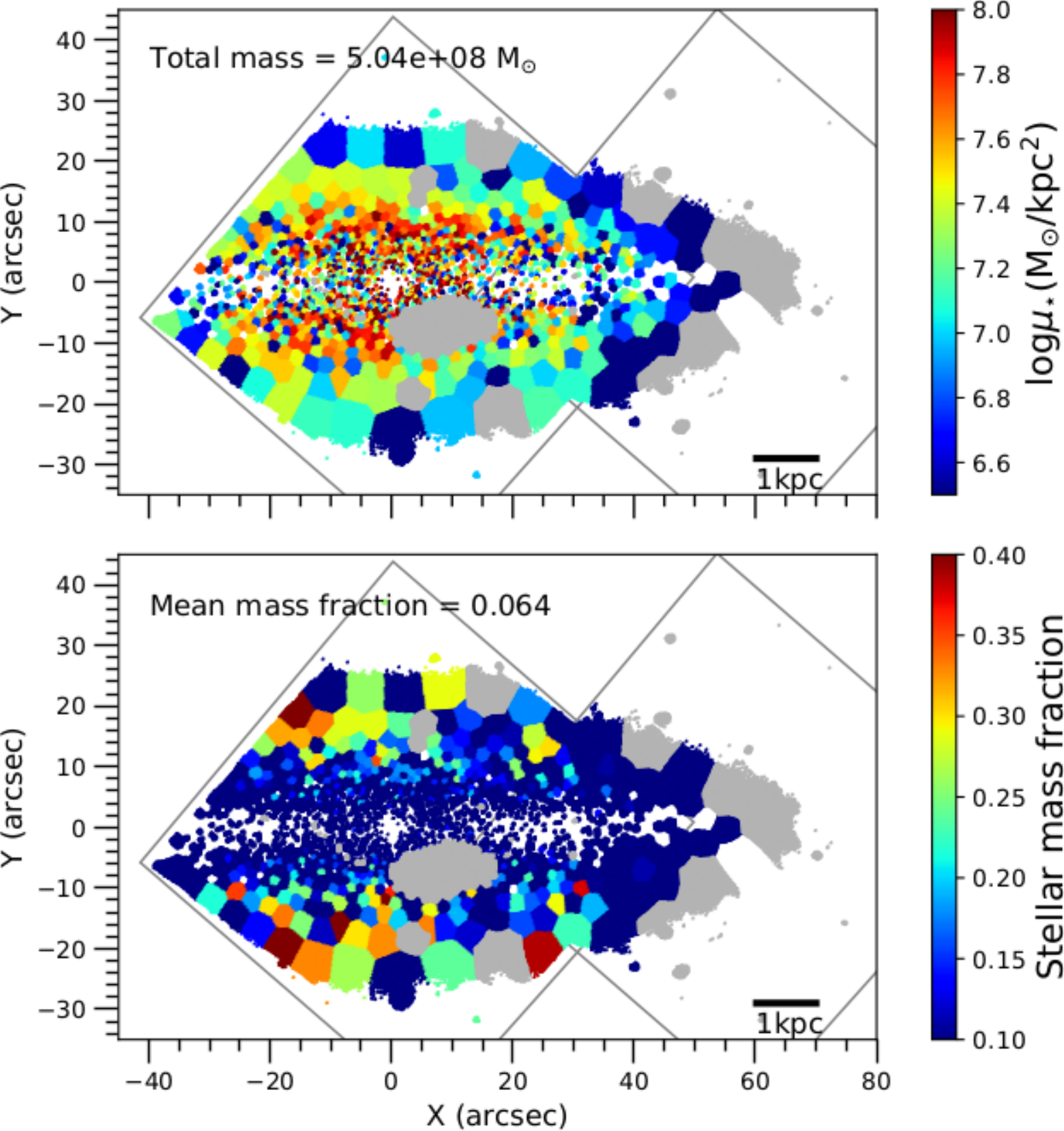}
}
\caption{As Fig.~\ref{fig:FCC153_accreted}, now for FCC\,177.
}
\label{fig:FCC177_accreted}
\end{figure*}

In addition, we mapped in Fig.~\ref{fig:FCC153_accreted} and \ref{fig:FCC177_accreted} the density distribution of the 
younger, metal-poor and $\alpha$-enhanced stars corresponding to the peak around 10\,Gyr in the SFHs of the thick disks.
In none of the two galaxies these populations are displayed in the dynamically coldest region of the thin disks (the closest to the midplane).
They are spread over the rest of both galaxies, mainly 
closer to the center (top panels).
The bottom panel of Fig.~\ref{fig:FCC153_accreted} indicates that in FCC\,153 the mass fraction of this population is more important at a certain height from the midplane.
In FCC\,177 (Fig.~\ref{fig:FCC177_accreted}), this is more clear and the mass of this population is more prominent in most bins of the inner thick disk than in the rest of the galaxy.
The estimates of the mass contribution by these younger populations, in the regions covered by our MUSE data of FCC\,153 and FCC\,177, are $\sim 10^9$\,M$_{\odot}$ and $\sim 5\times 10^8$\,M$_{\odot}$ respectively. \looseness-2

\section{Discussion}
\label{sec:discussion}
\subsection{Comparison to previous results} \label{sec:comparison}
FCC\,153 and FCC\,177 are not well-known galaxies, but previous long-slit studies revealed the stellar kinematics and the stellar populations along the major axis.
Our velocity maps in Figs.~\ref{chap4/fig:kin_FCC153} and \ref{chap4/fig:kin_FCC177} follow the rotation curves presented by \citet{Donofrio1995}. Nonetheless, we found lower values for $\sigma$ in both galaxies than what they showed in their paper. On the other hand, our kinematics are in very good agreement with \citet{Bedregal2006}, who attributed their discrepancy in $\sigma$ with \citet{Donofrio1995} to a resolution issue in the previous work.
\citet{Bedregal2006} also found no peculiarities in $h_3$ and $h_4$, as in our Figs.~\ref{chap4/fig:kin_FCC153} and \ref{chap4/fig:kin_FCC177}, confirming the simple velocity structure shown by the mean velocity map. 
Stellar populations extracted by \citet{Spolaor2010b}, \citet{Koleva2011} and \citet{Johnston2012} 
give slightly younger luminosity-weighted ages than our mass-weighted results in the midplane of FCC\,153 and FCC\,177, compatible with them. 
Our stellar kinematics and populations are also consistent with the recently published long-slit analysis performed by \citet{Katkov2019}.
\looseness-2

\subsection{Formation and evolution of S0 galaxies in the Fornax cluster} \label{sec:fornax}
In Paper~I and this paper, a set of three S0 galaxies within the virial radius of the Fornax cluster were analyzed.
In Paper~I, an overall old S0 was described, with no clear age distinction between the thin and the thick disk.
In this paper, we analyzed two galaxies quite similar to each other, but very different from FCC\,170.
In fact, they show a clear age distinction between the thin and the thick disk, 
similarly to the Milky Way and some external galaxies \citep[e.g.,][]{Gilmore1983,Haywood2015,Yoachim2008b,Comeron2015}.
This age difference is
sharper in FCC\,153, but clear in both galaxies suggesting a different formation process for the two geometrically defined disks.
While their thick disks formed faster during the early stages of the life of these two galaxies, from the initial metal-poor gas, 
the beginning of the thin-disk formation was probably slightly delayed (although we cannot see this small age difference in our SFHs).
This was discussed for FCC\,170 in Paper~I and would explain
 the presence in the thin-disk regions of old stars with higher metallicities and lower [Mg/Fe] abundance than in the thick disk (especially in FCC\,153, which probably initially evolved in a shorter timescale than FCC\,177).
In FCC\,153 and FCC\,177, the bulk of thin-disk stars would have formed slowly over an extended time. \looseness-2

Our stellar-population analysis points clearly to different evolutions for FCC\,170 from FCC\,153 and FCC\,177.
First of all, they have different total masses.
FCC\,170 is the most massive, followed by FCC\,153 and then FCC\,177, according to their maximum circular and rotation velocities.
\citet{Koleva2011} observed the down-sizing phenomenon in the Fornax cluster.
The star formation was shifted to galaxies with lower mass along time.
We observe this trend in our sample, since FCC\,177 has the lowest total mass and the youngest stars, while the most massive FCC\,170 hosts only old stars.
\citet{Kuntschner2000} confirmed a correlation between luminosity and age in lenticulars in Fornax, previously found in Coma \citep{Terlevich1999}. 
This means that, in the Fornax cluster, the systems hosting younger populations are predominantly fainter (S0) galaxies.
In addition, \citet{Kuntschner2000} and \citet{Bedregal2008} recovered some 
stellar-population properties, usually typical of ellipticals, in the central regions of brightest lenticulars in Fornax, including FCC\,170.
They proposed a faster formation timescale, than in the rest of lenticulars, for the most massive S0 galaxies, which are the oldest and most $\alpha$-enhanced.
This pointed towards different formation scenarios for bright and faint S0 galaxies. The authors suggested that the former (at least their central regions) would have a similar origin to ellipticals, since they have similar SFHs, while the latter would be more likely the results of spirals after loosing their gas.
The properties of FCC\,170 from our analysis correspond to the description of bright S0s.
FCC\,153 and FCC\,177, defined as "intermediate" (brightness) S0s by \citet{Bedregal2011}, already harbor young populations and their global SFH is noticeably different from FCC\,170. \looseness-2

As indicated in Fig.~\ref{chap4/fig:fornax} and Table~\ref{chap4/table:gal_prop}, the three galaxies are located at different distances from the cluster central galaxy FCC\,213: FCC\,170 is the closest, while FCC\,153 is the farthest.
FCC\,170 lives within the highest-density region of Fornax, where tidal interactions were proposed as the stripping mechanism traced by the spatial distribution of intracluster light \citep{Iodice2017a} and blue globular clusters \citep{dAbrusco2016,Cantiello2018}.
Moreover, as suggested in Paper~I (see also \citealt{Iodice2019}), 
FCC\,170 could have belonged to a primordial subgroup, before falling into the Fornax cluster.
There, its star formation could have been accelerated, forming a large mass in a short time.
The rest of gas (if any) would have been later stripped out causing an early quenching.
The higher mass of FCC\,170 (compared to the other two galaxies) and the presence of a massive bar in it, not seen prominently in FCC\,153 and FCC\,177, could also be related to its past environment different from the other two galaxies.
For instance, bar formation can be triggered by gravitational interactions \citep[e.g.,][]{Martinez2017} and might have happened during the preprocessing epoch in FCC\,170.
The fact that FCC\,153 and FCC\,177 are located in a lower-density intracluster medium,
in addition to the potential lack of a previous-subgroup phase,
probably caused the difference in the SFH of their thin disks.
Unlike FCC\,170, they could form stars over an extended period of time, until a few Gyr ago.
According to \citet{Kuntschner2000}, young S0s could have been accreted recently to the Fornax cluster, since they tend to be in the periphery. \looseness-2

Although FCC\,153 and FCC\,177 display common properties in contrast to FCC\,170, they also show some individual peculiarities.
\citet{Iodice2019} studied the photometric properties of early-type galaxies in the Fornax cluster, in relation to their locations.
They found a clear correlation of galaxy morphology and color with environmental density. Red galaxies are distributed in a wide region where X-ray emission, indicating high hot-gas density, was detected \citep{Paolillo2002}. Bluer galaxies are located farther from the center.
The three galaxies in our sample are not located in regions with the same intracluster density and display different colors.
FCC\,177 and FCC\,170 are redder and located within the region traced by the X-ray emission.
FCC\,153 resides still within the cluster virial radius, but lives in a lower intracluster density region and is bluer.
Some properties of the thin disks could be related to this segregation.
FCC\,153's thin disk is very thin and its chemical properties are spatially well defined.
In FCC\,177 it is thicker and fuzzier. Some mixing between stars with slightly different stellar-population properties might be suggesting disk-heating processes were effective in red galaxies in their denser environment. \looseness-2

\citet{Iodice2019} also found thin-disk flaring structures in the photometric analysis of the three galaxies. 
The thick flares have similar morphologies in FCC\,170 and FCC\,153, although the former's one appeared more extended. In contrast, FCC\,177 showed a thinner flaring structure.
However, only the flares of FCC\,153 and FCC\,177 were found to be prominent in $g-i$. 
Their colors are different and could be suggesting different flaring-formation mechanisms, related to the different locations of the two galaxies.
FCC\,153 displays bluer colors in its outskirts, while FCC\,177 is redder especially in the flaring region. 
\looseness-2

In our stellar-population maps, the best spatially defined flare is found in FCC\,153, where the region including flaring stellar populations is clearly distinguished from the rest.
Especially, in the metallicity and velocity-dispersion maps,
flares of progressively more metal-rich and dynamically-colder populations are sharply evident from a certain radius (about 2~kpc) towards the outskirts.
In the metallicity map, the morphology of this flare suggests that it could be composed by different "monometallicity" flares. 
This means that stars with different metallicity would flare at different distances from the galactic center.
The flare is slightly more confused in [Mg/Fe] and only hinted in age.
\looseness-2

The stellar-population properties of the flares in FCC\,170 and FCC\,177, on the contrary, look more mixed and no age radial gradient is hinted.
This could be another indicator of the different flare origins suggested by the colors from \citet{Iodice2019} (the flare is bluer in FCC\,153).
For FCC\,170 and FCC\,177 gravitational interactions could have dynamically heated the thin-disk stars forming the observed more "disordered" flares. The composition of FCC\,153's flare by several monometallicity flares could be suggesting a secular-evolution origin.
However, although spatial bins in the flare display younger ages, we do not see in FCC\,153 the age radial gradient predicted by the inside-out formation proposed by \citet{Minchev2015}. Instead, the striking anticorrelation between metallicity and velocity dispersion seems to be suggesting a flaring of dynamical origin \citep{Narayan2002a,Narayan2002b}.\looseness-2

\subsection{Thick disks} \label{thick}
The thick disks in FCC\,153 and FCC\,177 could have been formed in the same two-step process proposed for FCC\,170 in Paper~I.
In a first phase, $2-4$\,Gyr long, most of the thick-disk stars were quickly formed in situ. This was followed by a second phase when a minor but significant stellar component was accreted from one or more satellites. 
The oldest in-situ populations, given their chemical composition in general so different from the thin disk (Section~\ref{sec:chem}), could have been formed already as a thick component, on a fast timescale, from turbulent gas. This fast first phase still allowed some chemical evolution (visible in Figs.~\ref{chap4/fig:agemet_hist} and \ref{chap4/fig:agealpha_hist}) until the second ex-situ components came up. 
However, an alternative origin could be possible, especially for FCC\,177.
Here, dynamically colder stars extend above $z_{\text{c1}}$. The transition between thin and thick-disk populations extends over a thick portion of the observed region and is not as clear as in FCC\,170 and especially FCC\,153. 
Because of the continuity in these properties between the thin and the thick disk, we cannot rule out disk heating or upside-down formation of a unique (thick) disk component in FCC\,177. \looseness-2

The younger ex-situ populations have roughly similar properties in the three thick disks (seen as bumps in the SFHs), 
in spite of being hosted by galaxies with different properties and in different locations of the cluster.
They show mean ages between 8 and 11\,Gyr, [M/H]\,$\sim -0.6$ and [Mg/Fe]\,$\sim 0.34-0.36$.
We consider accretion the best possibility to explain that these additional components formed later and with their peculiar chemical properties, when the rest of the host galaxies were already forming more chemically enriched stars.
Accreted satellites (one or more than one per galaxy) 
would have a lower mass than the host galaxy and hence a lower metallicity. 
In proportion to the host galaxy mass, the satellite(s) 
would have contributed similar mass fractions in the three galaxies, for the selected levels of regularization.
In FCC\,153 and FCC\,170, this fraction would be about 4\,\% of the total stellar mass, and 5\,\% in FCC\,177. In the respective thick disks, the contribution of the accreted populations would be higher. As measured in the region covered by our data, it would correspond to 17\,\% of the thick-disk total mass for FCC\,177 and FCC\,170 and 23\,\% for FCC\,153.
Spavone et al. (in preparation) suggested that accretion contributed considerably to the mass assembly of these three galaxies. Fractions of accreted mass were estimated from the photometric properties of their stellar halos (using the method detailed in \citealt{Spavone2017,Spavone2018}), leading to much larger values (especially for FCC\,170 and FCC\,177) than what we associate here to the younger-metal-poor component. In fact, their fractions include also the populations with the same properties as the in-situ component, that are therefore indistinguishable with our method. 
\looseness-2

We have checked what stellar mass these satellites should have had, corresponding to their [Fe/H], according to the stellar mass-metallicity relation for Local Group dwarfs, in Fig.~9 of \citet{Kirby2013}.
Using our approximate mean values of [M/H] and [Mg/Fe] to calculate $\text{[Fe/H] = } \text{[M/H]} - 0.75 \text{[Mg/Fe]}$ \citep{Vazdekis2015}, we estimated a stellar mass of about 10$^9$\,M$_{\odot}$ for the satellites.
We should be cautious in associating these accreted masses to the initial satellite masses.
These galaxies, before being accreted, could have had a total mass rather larger than the contribution that we see now in their hosts. They could have lost a significant mass fraction, still at large distances from the host galaxy, during the early stages of their accretion.
Although we discourage a quantitative comparison to other studies based on absolute values (see Discussion in Paper~I),
our values of [Mg/Fe] are in general higher than what was  expected from previous works on galaxies of about $10^9$\,M$_{\odot}$ with comparable total metallicities (e.g., the Large Magellanic Cloud, \citealt{vanderSwaelmen2013}, with a stellar mass of about $2\times 10^9$\,M$_{\odot}$, \citealt{Kim1998}).
A larger (initial) satellite mass could help to explain these chemical abundances, together with the dense environment where these satellites could have lived. \citet{Sybilska2017} showed as dwarfs and ellipticals in clusters tend to display higher [Mg/Fe] abundances than galaxies with similar stellar masses in the field (see also \citealt{Bernardi2006}). \looseness-2

In the three galaxies, the accreted stars were formed (in the satellite) before the host-galaxy quenching, as shown by the SFHs. For FCC\,170, not much time passed in between and we propose that these satellites were accreted soon after their formation. This would not give them time to be contaminated by supernovae type-Ia and could explain their $\alpha$-enhancement.
We suggest that they were accreted before cluster velocity dispersions were high enough to make direct mergers unlikely, probably before the infalling of these galaxies into the Fornax cluster.
It could have happened during an intense merger epoch around redshift $1-1.5$ \citep[e.g.,][]{Brook2004}. \looseness-2

The difference in metallicity and [Mg/Fe] abundance between the old (in-situ) and the young (ex-situ) populations is much sharper in the SFH of FCC\,170 than in FCC\,153 and FCC\,177 (Figs.~6 and 9 in Paper~I and Figs.~\ref{chap4/fig:agemet_hist} and \ref{chap4/fig:agealpha_hist} in this paper).
In the thick disks of the latter, the transition towards the most metal-poor and $\alpha$-enhanced youngest stars is rather gradual.
On the one hand, this smoothing could be (at least partially) real, given that our SFHs represent the mean properties of all populations of a certain age. Therefore, in the transition between the old and the younger components, we go (more or less gradually) from ages still dominated by the in-situ populations to the ex-situ-dominated populations. The two of them are partially overlapped in the SFH.
On the other hand, we should also consider this transition to be smoothed by the regularization used in pPXF. In general, this introduces a smoothing between different-population ages, metallicities and [Mg/Fe], probably reproducing the real behavior in most cases, but maybe sometimes introducing unreal or excessive smoothing, if the real transition between populations is sharp (as expected from accretion).
These two interpretations (regularization and overlapping of in-situ and ex-situ stars in the SFH) could explain also the apparent increase of [Mg/Fe] along time in the nuclear and thin disks and in the box/peanut of FCC\,170 (Paper~I). An alternative scenario could be the "wet" nature of the satellite merger, which would have accreted less evolved gas resetting the [Mg/Fe] "clock".
These and other interpretations were discussed for FCC\,170 in Paper~I. \looseness-2

Another explanation for the $\alpha$-enhancement in the thick disks and especially in their younger populations would be a varying initial mass function (IMF)  \citep{Martin2015a,Martin2015b,Labarbera2016,Lyubenova2016,Martin2016}.
Historically, the IMF has been thought to be universal, similar for systems with different properties \citep[e.g.][]{Kroupa2002a,Bastian2010}.
However,
numerous recent studies have proposed a non-universal IMF to solve different dilemmas, including
the excess of low-mass stars and the unrealistically short formation timescale derived from [Mg/Fe] abundances in early-type galaxies \citep[e.g.][]{Kuntschner2000,Treu2010,vanDokkum2010,Spiniello2015,Martin2015a,Martin2016,Lyubenova2016}.
If the IMF is "top heavy" (i.e. massive stars have more weight), more supernovae type II can eject $\alpha$-elements to the interstellar medium. Therefore, a top-heavy IMF can also enhance [$\alpha$/Fe], with no need to invoke a faster timescale. 

The IMF slope may vary as well in time. 
First generations would have been dominated by massive stars (flatter IMF) and following generations by low-mass stars (steeper slope) \citep[e.g.][]{Vazdekis1996,Martin2016}.
In our sample, time-varying IMF slope would explain the chemical evolution in individual regions without invoking a timescale change.
In a similar way, a flatter IMF in the potentially accreted satellites than in the host galaxy could justify their chemical composition.
Furthermore the IMF, interpreted as a galaxy local property rather than global \citep{Martin2015a,Labarbera2016,vanDokkum2017}, could have been different, for example, in the thick disk.
Ongoing studies on the IMF in the Fornax\,3D galaxies will probably provide helpful insights to this discussion. \looseness-2

\subsection{NSCs in the Fornax cluster}
NSCs were detected in the three galaxies of our sample by \citet{Turner2012}.
Their blue colors are in agreement with the young ages we found in the central regions of FCC\,153 and FCC\,177.
However, given their reduced size and the prominent boxy bulge in the center of FCC\,170, the young nucleus of this galaxy was not visible in our maps. \looseness-2

The two most invoked scenarios to form NSCs are the infall of a star cluster towards the galaxy center, due to dynamical friction, and the inflow of gas to the nuclear region, triggering star formation in situ.
Recent works by \citet{Kacharov2018} and \citet{Georgiev2019} have shown that different scenarios could have played a major role in different galaxies, but also that both of them could have provided different contributions to the same NSC.
Lyubenova et al. (submitted) supported cluster merging in a study including also FCC\,177 and FCC\,170. \looseness-2

The bursty SFH measured in our stellar-population analysis fits well with different phases of gas accretion.
However, we cannot rule out cluster merging as additional scenario, since it would have contributed populations very difficult to distinguish from the rest of contributions in the line of sight.
The gas could have come from mergers, supernova winds or disk instabilities \citep[e.g.,][]{Ordenes2018}.
\citet{Kuntschner2000} suggested that galaxy harassment could have driven some disk gas towards the galaxy center of young S0s in the Fornax cluster, causing a final burst. The fact that, in our plots from Fig.~\ref{chap4/fig:sfh} to \ref{chap4/fig:agealpha_hist}, the final bursts happen in the nuclei when the star formation is being quenched in the respective thin disks points to this scenario.
Following the discussion in \S~\ref{thick}, the low [Mg/Fe] since the formation of the first stars in the nuclei could be the hint of a variable IMF.
We speculate that if the gas was accreted in different stages in low-mass amounts and immediately formed stars, exhausting the gas until a new accretion, massive stars were never abundant enough to enhance [Mg/Fe]. 
\looseness-2

\section{Summary and conclusions}
\label{sec:conclusions}
This work completes the study of thick disks in the three edge-on S0 galaxies of the Fornax cluster, started in Paper~I. 
FCC\,153 and FCC\,177, located in a less dense intracluster medium compared to FCC\,170, appear very different from it, indicating a different evolution history.
Unlike FCC\,170, the stellar kinematics and populations of FCC\,153 and FCC\,177 reveal structures composed of relatively young thin disks embedded in old thick disks.
A clear spatial correspondence exists between mean stellar age and chemical properties. Youngest stars are the most metal-rich and least $\alpha$-enhanced, concentrated in the mid plane and the very central region. The thick disks are the most metal poor and $\alpha$-enhanced components.

Both galaxies display different bursts of star formation in their nuclei, until very recent times, suggesting a central growth driven by gas infall in different events.
Thin disks formed their mass in an extended time and did not suffer from an early quenching as in FCC\,170.
FCC\,153's thin disk appears very thin and geometrically well defined, with a flaring clearly seen in velocity dispersion, [Mg/Fe] and especially metallicity. Radial gradients are seen in these properties both in the thin and thick disks.
FCC\,177 has a more fuzzy and thicker thin disk. The transition towards the thick disk is less sharp than in FCC\,153 and the thin-disk flaring is not obvious.
The differences between the two galaxies could be related to the different density in their locations. While FCC\,177 and FCC\,170 live in the red-galaxy higher-density area traced by the X-ray emission, FCC\,153 is bluer and is located outside of this hot dense gas.

Thick-disk SFHs fit the same two-phase formation scenario that we proposed for FCC\,170. In a first in-situ phase the dominant old component would be formed. In the later phase a second younger component, significant but minor and formed ex situ with different chemical properties, would have been accreted.
Since our study on the origin of thick disks in external galaxies is so far limited to three lenticular galaxies located in the same cluster, a more extended analysis will soon include late-type galaxies in the field.

\begin{acknowledgements}
This work is based on observations collected at the European Organization for Astronomical Research in the Southern Hemisphere under ESO programme 296.B-5054(A).
FP acknowledges Fundaci\'on La Caixa for the financial support received in the form of a Ph.D. and post-doc contract. FP, JFB and GvdV and RL 
acknowledge support from grant 
AYA2016-77237-C3-1-P from the Spanish Ministry of Economy and Competitiveness 
(MINECO).
EMC and LM acknowledge financial support from Padua University through grants DOR1715817/17, DOR1885254/18 and BIRD164402/16.
RL acknowledges funding from the Natural Sciences and Engineering Research Council of Canada PDF award.
GvdV acknowledges funding from the European Research Council (ERC) under the European Union's Horizon 2020 research and innovation programme under grant agreement No 724857 (Consolidator Grant ArcheoDyn).
\end{acknowledgements}





\bibliographystyle{aa}
\bibliography{biblio}
\appendix
\section{Two-dimensional view of the SFH}
\begin{figure*}
\centering
\resizebox{.72\textwidth}{!}
{
\includegraphics[scale=1,width=\textwidth]{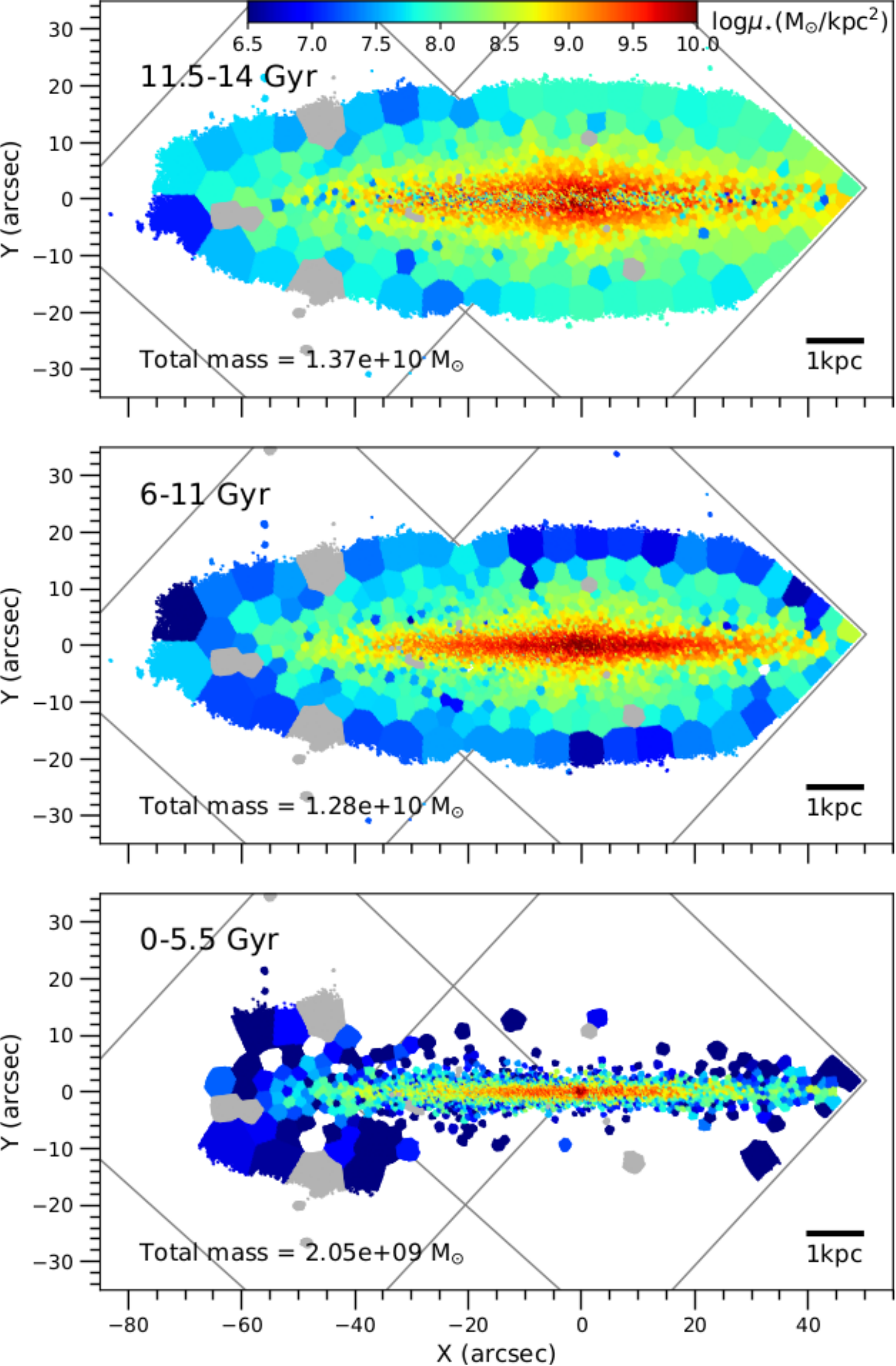}}
\caption{Two-dimensional view of the star formation history in FCC\,153. 
Stellar populations are mapped according to their age. Each map corresponds to a specific age bin: 
11.5\,--\,14\,Gyr in the top panel, 6\,--\,11\,Gyr in the middle panel and 0\,--\,5.5\,Gyr in the bottom panel. 
The color scale shows the mass density corresponding to the populations in the specific age 
bin.
The total mass in the age bin is indicated on bottom-left of each map.
The coverage of the two MUSE pointings and the discarded spatial bins are plotted in grey. A scale bar on bottom-right
of each map indicates the correspondence with physical units.
}
\label{appA/fig:age_bins_FCC153}
\end{figure*}

\begin{figure*}
\centering
\resizebox{.58\textwidth}{!}
{
\includegraphics[scale=1,width=\textwidth]{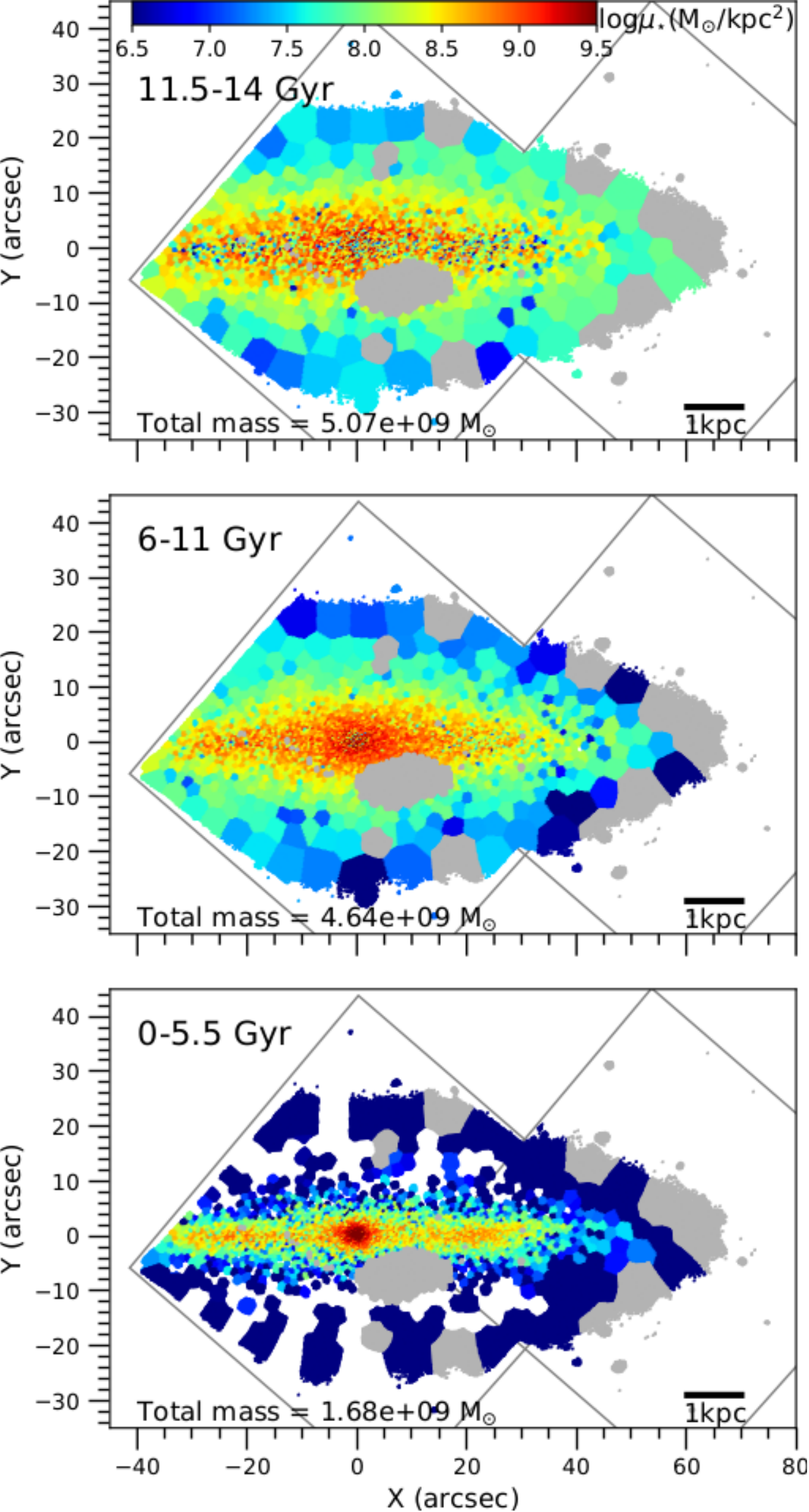}}
\caption{As Fig.~\ref{appA/fig:age_bins_FCC153}, now for FCC\,177. 
}
\label{appA/fig:age_bins_FCC177}
\end{figure*}

A two-dimensional view of the SFH is given by the maps in Figs.~\ref{appA/fig:age_bins_FCC153} and  \ref{appA/fig:age_bins_FCC177}.
Here, the stellar populations are divided into three age bins: 11.5\,--\,14\,Gyr, 6\,--\,11\,Gyr and 0\,--\,5.5\,Gyr. 
In each map, we have color coded the mass density contained in the corresponding age bin.
Populations more than 11.5\,Gyr-old (top panels) contribute about 47\,\% in FCC\,153 and 46\,\% in FCC177. 
In FCC\,153 they are less dense right in the midplane (than above and below) but aligned in the central region of the thin disk.
The $\sim$\,10\,Gyr-old populations are plotted in the middle-panel maps, including also populations in a broader age range, mainly concentrated in the thin disks.
In FCC\,153 (Fig.~\ref{appA/fig:age_bins_FCC153}), stars younger than 5.5\,Gyr are present only in the dynamically coldest region of the thin disk, but with a flaring in the outer part. In FCC\,177 (Fig.~\ref{appA/fig:age_bins_FCC177}), these young stars are distributed in a different shape and mainly in the nuclear star cluster region. \looseness-2

\section{Spatial distribution of metallicity in different ranges}
\begin{figure*}
\centering
\resizebox{.72\textwidth}{!}
{
\includegraphics[scale=1,width=\textwidth]{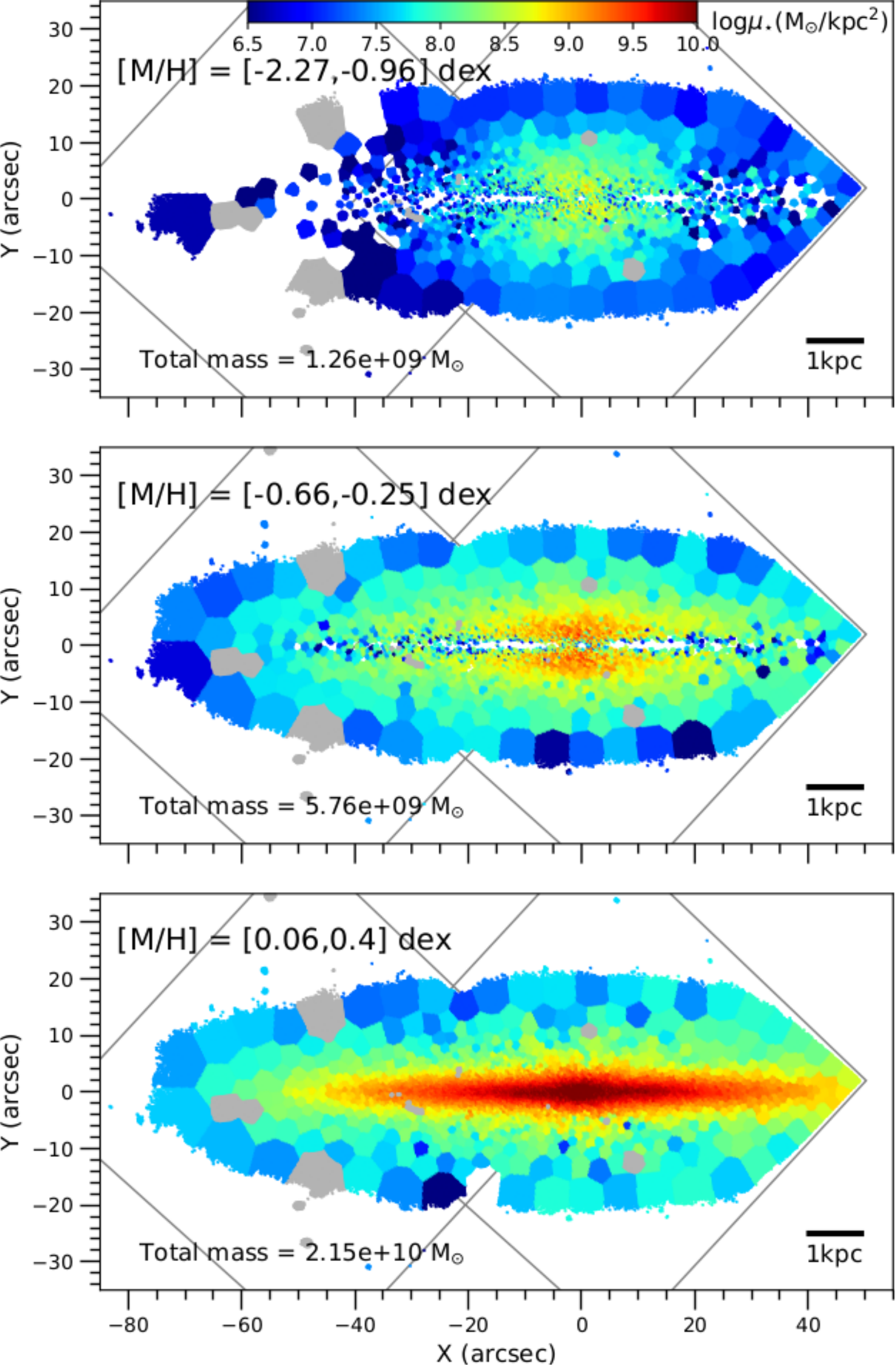}}
\caption{Maps of stellar populations in FCC\,153 with metallicities in three different bins: [M/H]\,=\,[-2.27,-0.96]\,dex in the top 
panel,
[-0.66,-0.25]\,dex in the middle panel and [0.06,0.4]\,dex in the bottom panel. 
The color scale shows the mass density corresponding to the populations in the specific metallicity 
bin.
The total mass in the metallicity bin is indicated on bottom-left of each map.
The coverage of the two MUSE pointings and the discarded spatial bins are plotted in grey. A scale bar on bottom-right
of each map indicates the correspondence with physical units.
}
\label{appA/fig:met_bins_FCC153}
\end{figure*}

\begin{figure*}
\centering
\resizebox{.58\textwidth}{!}
{
\includegraphics[scale=1,width=\textwidth]{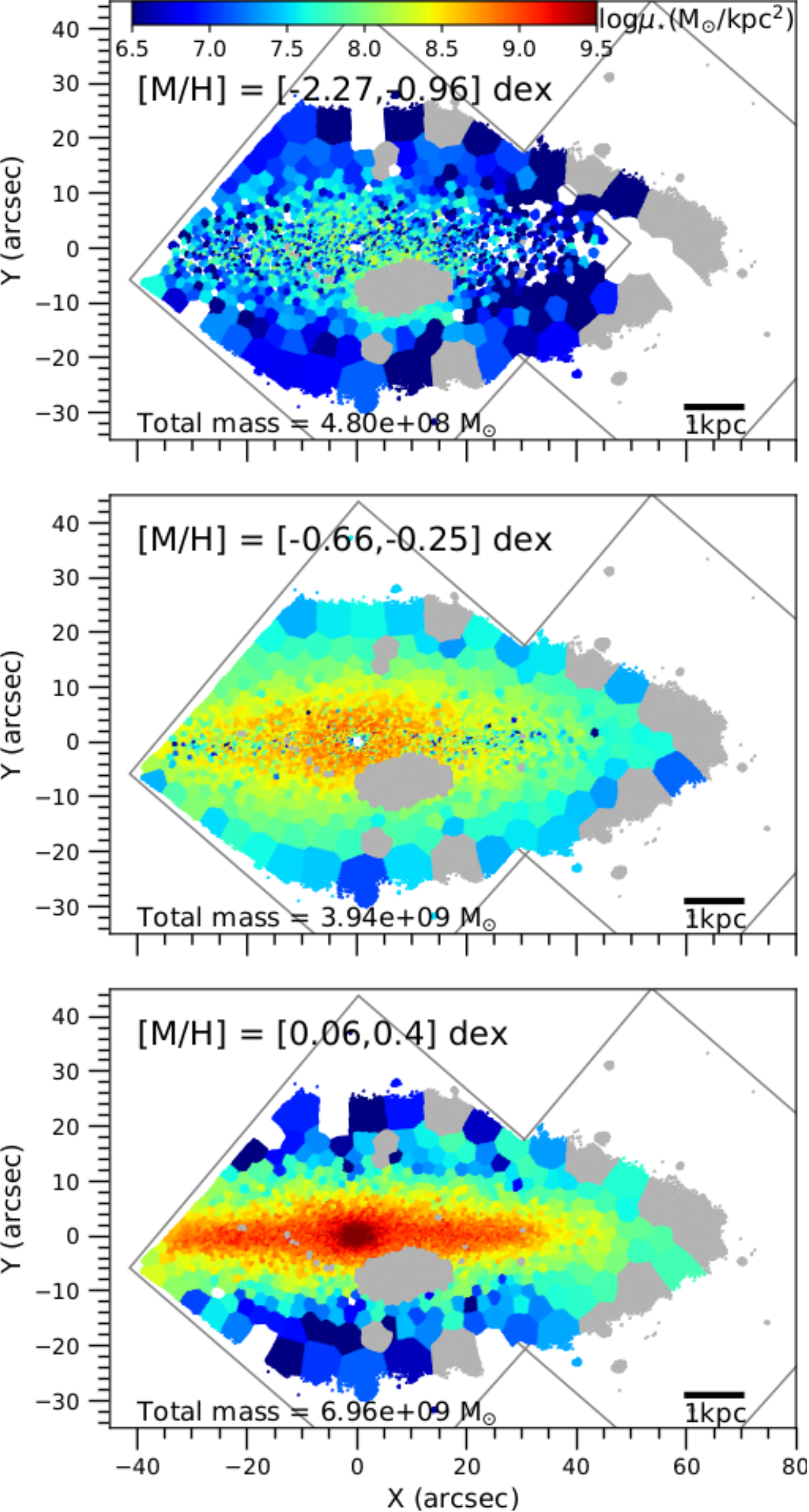}}
\caption{As Fig.~\ref{appA/fig:met_bins_FCC153}, now for FCC\,177.
}
\label{appA/fig:met_bins_FCC177}
\end{figure*}

Dividing our metallicity range into three bins, we can map the mass density of populations with different metallicities in our three galaxies.
The top panel of Figs.~\ref{appA/fig:met_bins_FCC153} and \ref{appA/fig:met_bins_FCC177} shows the most metal poor stars.
These stars give the lowest contribution to the total mass: about 4\,\% in the two galaxies. They are absent in numerous spatial bins of the outer regions of the two galaxies.
Stars with subsolar metallicities (top and middle panels) are present everywhere but are much denser towards the central region.
In FCC\,153, they are absent in a narrow band lying on the midplane.
Populations in the supersolar metallicity range (bottom panels) contribute most of the mass: about 74\,\% in FCC\,153 and 63\,\% in FCC\,177. 
They follow the shape of the thin disks, dominated by these metal-rich stars, in the three galaxies.
The central mass concentration is also completely dominated by these most metal-rich stars in FCC\,177 (including the boxy bulge and the NSC). \looseness-2
\end{document}